\documentclass[12pt]{article}

\usepackage{epsf,amsfonts,hyperref}
\renewcommand{\appendix}[1]{
    \addtocounter{section}{1}
    \setcounter{equation}{0}
    \renewcommand{\thesection}{\Alph{section}}
    \section*{Appendix \thesection\protect\indent #1}
    \addcontentsline{toc}{section}{Appendix \thesection\ \ \ #1}
}
\newcommand\encadremath[1]{\vbox{\hrule\hbox{\vrule\kern8pt 
\vbox{\kern8pt \hbox{$\displaystyle #1$}\kern8pt} 
\kern8pt\vrule}\hrule}}
\def\enca#1{\vbox{\hrule\hbox{
\vrule\kern8pt\vbox{\kern8pt \hbox{$\displaystyle #1$}
\kern8pt} \kern8pt\vrule}\hrule}}

\newcommand\figureframex[3]{
\begin{figure}[bth]
\hrule\hbox{\vrule\kern8pt 
\vbox{\kern8pt \vbox{
\begin{center}
{\mbox{\epsfxsize=#1.truecm\epsfbox{#2}}}
\end{center}
\caption{#3}
}\kern8pt} 
\kern8pt\vrule}\hrule
\end{figure}
}
\newcommand\figureframey[3]{
\begin{figure}[bth]
\hrule\hbox{\vrule\kern8pt 
\vbox{\kern8pt \vbox{
\begin{center}
{\mbox{\epsfysize=#1.truecm\epsfbox{#2}}}
\end{center}
\caption{#3}
}\kern8pt} 
\kern8pt\vrule}\hrule
\end{figure}
}

\newcommand{\eq}[1]{Eq.~(\ref{#1})}

\newcommand{\beq}{\begin{equation}}
\newcommand{\eeq}{\end{equation}}
\newcommand{\bea}{\begin{eqnarray}}
\newcommand{\eea}{\end{eqnarray}}

\renewcommand{\and}{{\qquad {\rm and} \qquad}}

\newcommand{\virg}{{\qquad , \qquad}}

 \newcommand{\Tr}{{\,\rm Tr}\:}

\newcommand{\td}[1]{{\tilde{#1}}}

\renewcommand{\l}{\lambda}
\newcommand{\om}{\omega}

\newcommand{\ee}[1]{{{\rm e}^{#1}}}

\newcommand{\Pint}{{\int\kern -1.em -\kern-.25em}}

\renewcommand{\Re}{{\mathrm{Re}}}

\newcommand{\Perm}{{\Sigma}}

\renewcommand{\l}{\lambda}

\newcommand{\curve}{{\cal E}}

\begin{document}
\sloppy


\pagestyle{empty}
\hfill SPT-08/056
\addtolength{\baselineskip}{0.20\baselineskip}
\begin{center}
\vspace{26pt}
{\large \bf {All orders asymptotic expansion of large partitions}}
\newline
\vspace{26pt}

{\sl B.\ Eynard}\hspace*{0.05cm}\footnote{ E-mail: eynard@spht.saclay.cea.fr }\\
\vspace{6pt}
Institut de Physique Th\'{e}orique de Saclay,\\
F-91191 Gif-sur-Yvette Cedex, France.\\
\end{center}

\vspace{20pt}
\begin{center}
{\bf Abstract}
The generating function which counts partitions with the Plancherel measure (and its q-deformed version), can be rewritten as a matrix integral, which allows to compute its asymptotic expansion to all orders.
There are applications in statistical physics of growing/melting crystals, T.A.S.E.P., and also in algebraic geometry.
In particular we compute the Gromov-Witten invariants of the $X_p= O(p-2)\oplus O(-p)\to {\mathbb P}^1$ Calabi-Yau 3-fold, and we prove a conjecture of M. Mari\~ no, that the generating functions $F_g$ of Gromov--Witten invariants of $X_p$, come from a matrix model, and are the symplectic invariants of the mirror spectral curve.

\end{center}
%





\vspace{26pt}
\pagestyle{plain}
\setcounter{page}{1}


\section{Introduction: partitions}

Partitions are extremely useful in many areas of physics and mathematics.
A partition with at most $N$ rows, is an ordered sequence of $N$ non-negative integers:
\beq
\l=(\l_1\geq \l_2\geq \dots \geq \l_N\geq 0)
\eeq
$|\l| = \sum_i \l_i$ is called the weight of the partition $\l$.
The length of the partition $n(\l)$ is the number of strictly positive rows $n(\l)=\#\{\l_i>0\}$.

\medskip

It is also convenient to represent the partition rotated by $\pi/4$,
i.e. we denote:
\beq
h_i = \l_i-i+N
\virg
h_1>h_2>\dots>h_N \geq 0
\eeq
$$
{\mbox{\epsfxsize=4.truecm\epsfbox{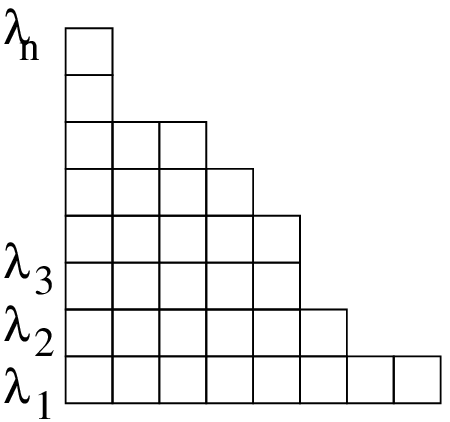}}}
\qquad
{\mbox{\epsfxsize=7.truecm\epsfbox{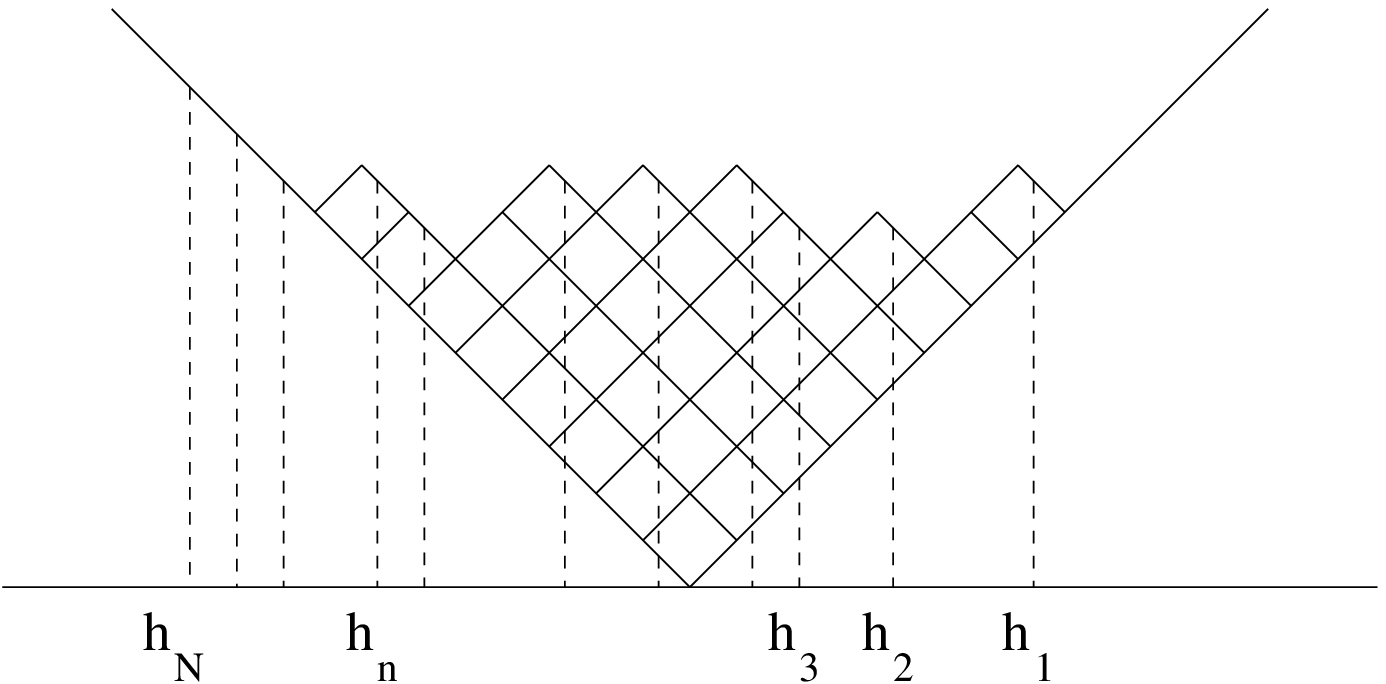}}}
$$
The $h_i$'s are in 1-1 correspondence with the downright edges of the rotated Ferrer diagram. 

\bigskip
Then, for various applications, ranging from statistical physics of growing/melting crystals \cite{spohn}, T.A.S.E.P. \cite{liggett, derrida}, to algebraic geometry \cite{Okounkov1},
one would like to measure partitions with the Plancherel measure ${\cal P}(\l)$ (described below), and compute partition functions of the form:
\beq
Z_N(q;t_k) = \sum_{n(\l)\leq N}\,\, {\cal P}(\l)\,\, q^{|\l|}\,\,\, \ee{-\sqrt{q}\,\sum_k {t_k\, q^{-k\over 2}\over k}\, C_k(\l)}
\eeq
where $C_k(\l)$ are the Casimirs of the partition $\l$ (see section \ref{Casimirs}).

Here in this article, we find {\bf all the coefficients} in the large $q$ expansion of $Z$:
\beq
\ln{\left(Z_N(q;t_k)\right)} = \sum_{g=0}^\infty\, q^{1-g}\,\, F_g(t_k) + {\rm exponentially\, small}
\eeq
$F_0$ and $F_1$ have been known for some time \cite{Okounkov2}, and here, we prove that all the other orders $F_g$'s with $g\geq 2$ are the symplectic invariants defined in \cite{EOFg}.

\bigskip

Beyond computing the partition function, one may also be interested in finding the large size asymptotic  shape of those random partitions.
The shape of the typical partition has been known for some time \cite{vershik, Planchrel1, Logan, kerov1, kerov2, kerov3, Okounkov1,Okounkov2}, it is related to Tracy Widom law \cite{TW}.
Here in this article, we recover the shape of the typical partition, and we compute {\bf corrections to all orders}. We also compute all orders expansions of density correlation functions.

\bigskip

We are also interested in a $q-$deformed version of the Plancherel measure ${\cal P}_q(\l)$, which has applications to algebraic geometry/topological string theory, in particular in computing the Gromov--Witten invariants of some Calabi--Yau 3-folds $X_p$ described in section \ref{secXp} below (see \cite{CGMPS, Marcosconj}).
Here in this article, we prove, as conjectured by M. Mari\~ no \cite{Marcosconj}, 
that the Gromov-Witten invariants of $X_p$ are given by a matrix model, and more precisely by the symplectic invariants of \cite{EOFg} computed for a spectral curve of the form $H(\ee{x},\ee{y})=0$.
This is typically the form of a mirror spectral curve in the type B topological string theory.
Moreover, using the recent result of \cite{DVKS}, this also proves that this model is equivalent to the Kodaira Spencer field theory.

\bigskip
{\noindent \bf Outline of the article:}

$\bullet$ Section 1 is an introduction, where we recall the definition of the Plancherel measure, the Casimirs.

$\bullet$ In Section 2, we rewrite the partition function $Z$ of the Plancherel measure, as a normal matrix integral.
As an immediate consequence, we find that the topological expansion of $\ln{Z}$ is given by the symplectic invariants of \cite{EOFg}, associated to a spectral curve which we compute explicitely.
This also gives all orders corrections to density correlation functions.

$\bullet$ In Section 3, we repeat the same derivation as in section 2, but for the $q$-deformed Plancherel measure partition function. We also rewrite it as a matrix integral, and as a consequence we find that its topological expansion, is again given by the symplectic invariants of \cite{EOFg}.
In particular, we study the consequences for the Gromov--Witten invariants of the Calabi-Yau toric 3-folds $X_p$. We prove the conjecture of \cite{Marcosconj} that the Gromov-Witten invariants, are the symplectic invariants of a mirror spectral curve.

$\bullet$ In section 4, we explain the main applications: crystal growth, T.A.S.E.P., length of the increasing subsequences of a random permutation, and algebraic geometry Gromov-Witten invariants.

$\bullet$ Section 5 is the conclusion.

\subsection{Plancherel measure}

The Plancherel measure is:
\beq
{\cal P}(\l) = \left({\rm dim}\,\l\over |\l|!\right)^2
= { \prod_{1\leq i<j\leq N} \,\,(h_i-h_j)^2\over \prod_{i=1}^N \,\,(h_i !)^2}
\eeq
where ${\rm dim}\,\l$ is the dimension of the representation of the symmetric group $\Sigma(|\l|)$, indexed by the partition $\l$.
The Plancherel measure depends only on the partition $\l$ and it does not depend on $N\geq n(\l)$.

Its $q-$deformed version is:
\beq
{\cal P}_q(\l) 
= {\prod_{i<j} [h_i-h_j]^2\over \prod_i ([h_i] !)^2}
\eeq
where $[h]$ is the $q$-number:
\beq
[h] = q^{-h/2}-q^{h/2}
\eeq
i.e.
\beq
{\cal P}_q(\l) 
= {\prod_{i<j} (q^{h_i}-q^{h_j})^2\over \prod_{i=1}^N q^{(N-1)h_i} \, q^{-{1\over 2}h_i(h_i-1)}\,\,\, \prod_{i=1}^N\prod_{j=1}^{h_i} (1-q^{j})^2 }
\eeq

\bigskip
The Plancherel measure appears in many physical and mathematical problems.
It is the natural measure on partitions.

Among famous physical problems related to Plancherel measure are the 2D growing/melting crystal \cite{Okounkov2, johansson}, the length of the longest increasing subsequence \cite{ssuites1}, or the Totally asymmetric exclusion process (T.A.S.E.P.) \cite{liggett, derrida}.
Plancherel measure also plays an important role in string theory and algebraic geometry, in the computation of Gromov-Witten invariants \cite{nekrasov, Okounkov1, Okounkov2, Marcosconj}.

\medskip
We develop those examples in section \ref{secapplis}.

\subsection{Expectation values and Casimirs}
\label{Casimirs}

In most applications, one would like to compute expectation values of the following form (as well as the $q$-deformed version of this):
\beq
Z_N(g_k) = \sum_{n(\l)\leq N}\,\, {\cal P}(\l)\,\,\,\prod_i \ee{- \sum_k g_k C_k(\l)}
\eeq
where $C_k(\l)$ is the $k^{\rm th}$ Casimir, which can be defined as the $k^{\rm th}$ term in the small $z$ expansion of:
\beq
\sum_k  {1\over k!} \, z^{k} C_k(\l) =  \sum_{i=1}^N \ee{z (h_i-N+{1\over 2})} + {\ee{-(N-{1\over 2})z}\over  \ee{z}-1}   - {1\over  z}
\eeq
The Casimirs depend only on $\l$, they don't depend on $N$.

We may write:
\beq
C_k(\l) = \sum_{i=1}^N (h_i-(N-{1\over 2}))^k\,\, + C_{k,0}(N)
\eeq

For instance:
\beq
C_1(\l)  = \sum_i h_i - {N(N-1)\over 2}-{1\over 24} =|\l|-{1\over 24}
\eeq
\beq
C_2(\l) 
= \sum_i h_i^2 -(2N-1)\sum_i h_i +{2\over 3}\,N (N-{1\over 2}) (N-1) 
= \sum_i \l_i (\l_i-2i+1)
\eeq
In general $\sum_k g_k C_k(\l)$ can be written:
\beq
\sum_k g_k C_k(\l) = \sum_k \td{t}_k \sum_i h_i^k = \sum_i A(h_i)
\eeq
and it is important to notice that the coefficients $\td{t}_k$ may depend on $N$.

\medskip 

Finally, what one would like to compute is a sum of the form:
\beq\label{defobservAh}
Z_N \propto \sum_{n(\l)\leq N}\,\,\, {\cal P}(\l)\,\,\,\prod_{i=1}^N \ee{- A(h_i)}
\virg
A(h) = \sum_{k=0}^{d+1} \td{t}_k h^k
\eeq

\vspace{1cm}

\section{Plancherel measure}

In this section, we rewrite the sum
\beq
Z_N(q,t_k) = \sum_{n(\l)\leq N}\, {\cal P}(\l)\,\,\, q^{|\l|}\,\, \ee{-\sum_{k=2}^{d+1} {{t_k\,q^{1-k\over 2}\over k}\,C_k(\l)}}
\eeq
as a matrix integral, and as a consequence, we obtain its topological large $q$ expansion:
\beq
\ln{Z_N(q,t_k)} \sim  \sum_{n=0}^\infty\, q^{1-n}\,\, F_n(t_k)
\eeq

\subsection{Transformation into a matrix integral}

Consider the contour ${\cal C}$ which goes above the positive Real axis in the negative direction and below the positive real axis in the positive direction, i.e. it encloses all non negative integers.
$$
{\mbox{\epsfxsize=10.truecm\epsfbox{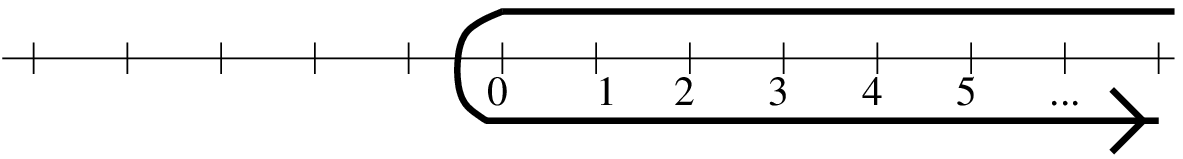}}}
$$

The function:
\beq
f(\xi) = -\xi\, \Gamma(-\xi)\Gamma(\xi)\,\,\ee{-i\pi \xi} = {\pi\,\ee{-i\pi \xi}\over \sin{(\pi \xi)}}
\eeq
has simple poles with residue $1$ at all integers ($\xi\in {\mathbf Z}$).

Therefore we have:
\bea
&& \oint_{{\cal C}^N} d\xi_1\dots d\xi_N\,\, \prod_{i<j} (\xi_i-\xi_j)^2\,\,\prod_{i} f(\xi_i)\,{\ee{- A(\xi_i)}\over \Gamma(\xi_i+1)^2}   \cr
&=& \sum_{h_1,\dots,h_N=0}^\infty \,\, \prod_{i<j} (h_i-h_j)^2\,\,\prod_{i} {\ee{- A(h_i)}\over h_i!^2}   \cr
&=& N!\,\sum_{h_1>\dots>h_N\geq 0} \,\, {\prod_{i<j} (h_i-h_j)^2\over h_i!^2}\,\,\prod_{i} \ee{- A(h_i)}   \cr
&=& N!\, \sum_{n(\l)\leq N} {\cal P}(\l)\,\, \prod_i \ee{-A(h_i)}
\eea

In other words, the Plancherel measure partition function is a normal\footnote{The ensemble $H_N({\cal C})$ is the set of normal matrices whose eigenvalues belong to ${\cal C}$, for instance $H_N({\mathbb R})=H_N$ is the set of hermitean matrices.} matrix integral:
\beq
\sum_{h_1>\dots>h_N\geq 0} \,\, {\prod_{i<j} (h_i-h_j)^2\over h_i!^2}\,\,\prod_{i} \ee{- A(h_i)}   
= {q^{N^2\over 2}\over N!}\,\int_{H_N({\cal C})} dM\, \ee{-\sqrt{q}\, \Tr V(M)}
\eeq
with the potential:
\beq
\sqrt{q}\, V(x) = \ln{\Gamma(\sqrt{q}\, x)}-\ln{\Gamma(-\sqrt{q}\, x)}+i\pi \sqrt{q}\, x +\ln{(\sqrt{q}\, x)} +  A(\sqrt{q}\, x) 
\eeq
$$
{\mbox{\epsfxsize=5.truecm\epsfbox{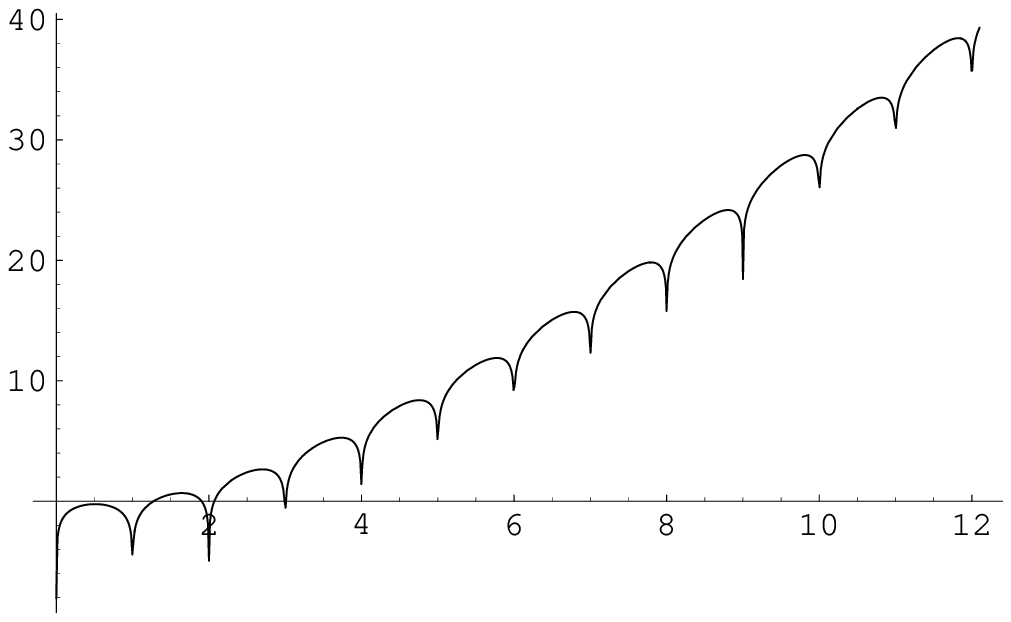}}}
$$

And we write:
\beq
A(\sqrt{q}\, x)  = \sqrt{q}\Big[ -x \ln{q} + \sum_{k=2}^{d+1} {t_k\over k}\,\left(x-{N-{1\over 2}\over \sqrt{q}}\right)^k + A_0 \Big]
=\sqrt{q}\, U(x) 
\eeq
Therefore, the derivative of the potential can be written (using Stirling's formula):
\bea
V'(x) 
&=& \psi(\sqrt{q} x)+\psi(-\sqrt{q} x) + i\pi + {1\over \sqrt{q} x} + U'(x)  \cr
&=& 2\ln{(\sqrt{q} x)}  +   U'(x) + {1\over \sqrt{q} x} - \sum_{n=1}^\infty {B_{2n}\over n (\sqrt{q} x)^{2n}} \cr
&=& 2\ln{x}  +  \sum_{k=1}^{d} t_{k+1}\,\left(x-{N-{1\over 2}\over \sqrt{q}}\right)^k + {1\over \sqrt{q} x} - \sum_{n=1}^\infty {B_{2n}\over n (\sqrt{q} x)^{2n}} \cr
\eea
where $B_n$ are the Bernouilli numbers, and $\psi=\Gamma'/\Gamma$.

\subsection{The rescaling factor and topological expansion}

We have introduced a rescaling factor $\sqrt{q}$ such that $h_i=\sqrt{q}\, x_i$, because typical $h_i$'s are not of order $1$, they are large, and with this rescaling, we expect the typical $x_i$'s to be of order $1$.

\smallskip

Another reason for introducing a rescaling factor $\sqrt{q}$, is that it will serve as an expansion parameter.

If the matrix model integral has a so-called "topological" large $q$ expansion \cite{thooft, BIPZ}, i.e. an expansion in powers of $q^{-1}$ here (and this must be the case by definition for formal series), then the coefficients are given by the solution of loop equations first found in \cite{eynloop1mat, CE}, and are given by the symplectic invariants \cite{EOFg}.
It was found in \cite{EOFg}, that the whole $1/q$ expansion can be written in terms of the spectral curve, i.e. the "large $q$ density":
\beq
\ln{Z_N} \sim \sum_{g=0}^\infty q^{1-g}\, F_g
\eeq
where $F_g$ are the symplectic invariants of the spectral curve.
Here, we shall see that the $F_g$'s don't depend on $N$ and on $q$.

\smallskip
Therefore in what follows we aim at determining the spectral curve corresponding to the potential $V(x)$ and to the integration contour ${\cal C}$.

\subsection{The spectral curve}\label{spcurverecipe}

The spectral curve is more or less the large $q$ density of the $h_i$'s. It is related to the  "large $q$ limit" of the resolvent $W(x)$ (see \cite{EOFg}):
\beq
W(x) = "{\rm lim}"\,\,\, {1\over \sqrt{q}}<\Tr {1\over x-M}> = "{\rm lim}"\,\,\,\left<\sum_i { 1\over \sqrt{q} x- h_i} \right>
\eeq
\bigskip 

Fixed filling fractions spectral curves can be found by the following recipe\footnote{The true definition of the spectral curve, is obtained by dropping the connected 2-point function from the loop equation, see \cite{EOFg}. But for 1-matrix models, this is more or less equivalent to solving the  saddle-point equations for the $h_i$'s, and this simplifies to give the recipe presented here.}:

we look for a set of cuts $[a_i,b_i]$ in the complex plane, and we look for a function $W(x)$ analytical outside the cuts, which behaves like ${N\over \sqrt{q}\, x}$ at $\infty$, and such that on each cut:
\beq\label{loopeqWxab}
W(x+i0)+W(x-i0)=V'(x) \virg \forall \, x\in [a_i,b_i]
\eeq
and
\beq
-{1\over 2i\pi} \oint_{[a_i,b_i]} W(x)dx =  \epsilon_i  
\eeq
where $\epsilon_i$ are the filling fractions.
Once the filling fractions are given, the solution is more or less unique.
Therefore, we have to specify filling fractions.

\bigskip

In fact, there are as many solutions of loop equations of matrix models (i.e. as many possible spectral curves corresponding to this potential), as the number of integration contours where the integral converges.
A choice of integration contour, is equivalent to a choice of filling fractions.

Here we want the integration contour to be ${\cal C}^N$, and we have to find which filling fractions it corresponds to.

\medskip
The spectral curve is the "large $q$" resolvent, i.e. it gives the asymptotic shape of a typical large partition.
For this problem, the typical shape of large partitions has been determined for a long time \cite{Planchrel1, Okounkov1, Okounkov2, nekrasov3}, and has a very special property called "arctic circle". We describe it below.


\subsubsection{Arctic circle}\label{secarcticcircle}

Our partition function $Z_N=\sum_{n(\l)\leq N} {\cal P}(\l) \, \ee{-A(\l)}$ depends on $N$ through the summation bound.
What was observed \cite{Jarctic}, is that typical partitions have a typical length $<n(\l)>=\overline{n}(q,t_k)$, and if we choose $N>\overline{n}$, the $h_i$'s with $N\geq i>\overline{n}$ (i.e. $h_i<N-\overline{n}$) of a typical large partition are "frozen", i.e. $h_i=N-i$, i.e. $\l_i=0$.

$\overline{n}$ is sometimes called the arctic circle, because everything above the arctic circle is frozen \cite{Jarctic}.
This means that asymptotically $Z_N \sim Z_{\overline{n}}$, up to exponentially small corrections, i.e. the $F_g$'s should not depend on $N$, and the $N$ dependence should be only in the exponentially small corrections:
\beq
\ln{Z_N} \sim \sum_{g=0}^\infty q^{1-g} F_g + O(\ee{-\sqrt{q}\, C(N)})
\eeq
This also means that, order by order in a $1/q$ expansion:
\beq
\ln{Z_\infty} \sim \ln{Z_N} \sim \ln{Z_{\overline{n}}} \sim \sum_g q^{1-g} F_g
\eeq
In other words, if the arctic circle property holds, we may determine the $F_g$'s by choosing $N=\overline{\nu}$ for instance, or by choosing our favorite value for $N$.
We can also try to find a spectral curve for any $N$, and then check that indeed the $F_g$'s are independent of $N$, and this is what we do below.

\subsubsection{Chiral ansatz}
\label{secchiralansatz}

We try to guess the form of the spectral curve, and then we have to check the consistency of our hypothesis.
\medskip

The fact that all the $h_i$'s beyond $\overline{n}$ are frozen means that the large $q$ resolvent (i.e. the spectral curve) can be written:
\beq\label{WtdWchiral}
W(x) = {1\over \sqrt{q}} \sum_{i={\overline{n}}+1}^{N} {1\over x-{N-i\over \sqrt{q}}}\,\,\, + \td{W}(x-b)
\virg
b={N-\overline{n}\over \sqrt{q}}
\eeq
where $\td{W}(x)$ has a cut $[0,a]$ with an edge at $0$.

\smallskip
In general, $\td{W}$ may have several cuts, depending on the $t_k$'s. Here we assume that the $t_k$'s are such that $\td{W}$ has only one cut. We discuss this asumption below in section \ref{secphasetr}.

\medskip

Thus, $W(x)$ corresponds to a multicut solution, with $N-\overline{n}+1$ cuts, one of them is $[b,b+a]$ with filling fraction ${\overline{n}\over \sqrt{q}}$, and the others are degenerate cuts $[{N-i\over \sqrt{q}},{N-i\over \sqrt{q}}]$ with filling fraction ${1\over \sqrt{q}}$.
This form of the spectral curve, is also called chiral ansatz in the litterature, after the solution of \cite{chiral, KSW}, see \cite{CGMPS}.

$\td{W}(x)$ is thus a one cut solution and behaves for large $x$ as
\beq
\td{W}(x) \sim {\overline{n}\over \sqrt{q}}\,{1\over x}
\eeq
and is such that
\beq\label{sdpeqWtd}
\td{W}(x+i0)+\td{W}(x-i0) = V'(x+b) - {2\over \sqrt{q}} \sum_{i={\overline{n}}+1}^{N} {1\over x+b-{N-i\over \sqrt{q}}} 
\qquad ,\forall x\in [0,a]
\eeq
In other words, we have to solve the same problem as for $W$, but with a new condition that the cut is of the form $[0,a]$.

Notice that:
\beq
{2\over \sqrt{q}} \sum_{i={\overline{n}}+1}^{N} {1\over x+b-{N-i\over \sqrt{q}}}
=  2\ln{x+b\over x} -{1\over \sqrt{q}x} + {1\over \sqrt{q}(x+b)}
+  \sum_{m=1}^\infty {B_{2m}\over m\, q^{m}}\,\, \left({1\over x^{2m}}-{1\over (x+b)^{2m}}\right)
\eeq
where $B_m$ are the Bernouilli numbers, and therefore the RHS of \eq{sdpeqWtd} is:
\bea
&& V'(x+b) - {2\over \sqrt{q}} \sum_{i={\overline{n}}+1}^{N} {1\over x+b-{N-i\over \sqrt{q}}} \cr
&=& 
2\ln{x}  +  \sum_{k=1}^{d} t_{k+1}\,\left(x-{\overline{n}-{1\over 2}\over \sqrt{q}}\right)^k 
 + {1\over \sqrt{q}x} 
-  \sum_{m=1}^\infty {B_{2m}\over m\, q^{m}\, x^{2m}}
\eea
Notice that it is independent of $N$.

\subsubsection{edge at zero}

We want to determine $\td{W}(x)$, which has only one cut $[0,a]$.

One cut solutions are better written in the Zhukovski parametrization:
\beq
x(z) = \gamma\,(z+2+{1\over z}) \virg a=4\gamma
\eeq
Notice that each value of $x$ corresponds to two values of $z$, namely $z$ and $1/z$.
We call the domain $|z|>1$ the physical sheet, and $|z|<1$ the second sheet.

Then we write:
\beq\label{defuknoq}
\sum_{k=1}^{d} t_{k+1}\,\left(x-{\overline{n}-{1\over 2}\over \sqrt{q}}\right)^k  
= \sum_{k=0}^d u_k (z^k+z^{-k})
\eeq
where the coefficients $u_k$'s are functions of $\gamma$.
We determine $\gamma$ by the condition:
\beq
\gamma=\ee{-u_0}
\eeq
The 1-cut resolvent $\td{W}(x)$ is then:
\beq
\om(z) = \td{W}(x(z)) 
= \sum_{k=1}^d u_k z^{-k}  + 2\ln{(1+1/z)}  +{1\over 2\sqrt{q} x(z)} - \sum_{n=1}^\infty {B_{2n}\over 2n (\sqrt{q} x(z))^{2n}}
\eeq
indeed it satisfies \eq{loopeqWxab}:
\beq
\om(z)+\om(1/z) =  2\ln{x}  +  \sum_{k=1}^{d} t_{k+1}\,\left(x-{\overline{n}-{1\over 2}\over \sqrt{q}}\right)^k 
 + {1\over \sqrt{q}x} 
-  \sum_{m=1}^\infty {B_{2m}\over m\, q^{m}\, x^{2m}}
\eeq
and it has (order by order in $1/\sqrt{q}$) no singularity in the physical sheet $|z|>1$, and it behaves like $O(1/z)$ for large $z$:
\beq
\om(z) \sim {\overline{n}\over \sqrt{q} x(z)} \sim  \left((u_1+2)\gamma+{1\over 2\sqrt{q}}\right)\,{1\over x(z)} + O (1/z^2)
\eeq
i.e. we find $\overline{n}$:
\beq
\overline{n} - {1\over 2} = (u_1+2)\gamma\,\,\sqrt{q}
\eeq
The $u_k$'s are then found from \eq{defuknoq}:
\beq\label{finduknoq}
\sum_{k=0}^d u_k (z^k+z^{-k})
= \sum_{k=1}^{d} t_{k+1}\,\ee{-k u_0}\, (z+{1\over z}-u_1)^k 
\eeq
In particular, the coefficient of $z^0$ and $z^1$ are:
\beq\label{equ0u1gen}
\left\{\begin{array}{l}
2u_0 = \sum_{k=1}^d (-1)^k\, t_{k+1}\,\ee{-k u_0}\, \sum_{j=0}^{[k/ 2]} {k!\, u_1^{k-2j}\over j! j! (k-2j)!} \cr \cr
u_1 = - \sum_{k=1}^d (-1)^k t_{k+1}\,\ee{-k u_0}\, \sum_{j=0}^{[(k-1)/2]} {k!\, u_1^{k-2j-1}\over j! (j+1)! (k-2j-1)!}
\end{array}\right.
\eeq
which means that (after eliminating $u_1$) $u_0$ satisfies a transcendental equation of the form
\beq
{\rm Polynomial}(u_0,\ee{-u_0})=0
\eeq
It is important to notice that this equation is independent of $N$ and $q$, and thus, $u_0, u_1$, and all the $u_k$'s depend only on the $t_k$'s, they don't depend on $N$ and $q$.

\bigskip

The resolvent $W(x)$ is thus given by \eq{WtdWchiral}:
\beq
W(x(z)) = {1\over \sqrt{q}} \sum_{i={\overline{n}}+1}^{N} {1\over x-{N-i\over \sqrt{q}}}\,\,\, + \om(z)
\virg
x(z)=b+\gamma(z+2+{1\over z})
\eeq

The spectral curve is the density $y={1\over 2}(W(x-i0)-W(x+i0))$, i.e. the discontinuity of $W$ along the cut, i.e. $y(z)={1\over 2}(\om(1/z)-\om(z))$, therefore our spectral curve is:
\beq\label{spcurvenoq}
\encadremath{
{\cal E}(t_k) =
\left\{\begin{array}{l}
\displaystyle x(z) = {N-{1\over 2}\over \sqrt{q}}+\ee{-u_0}\,(z+{1\over z}-u_1) \cr \cr
y(z) = \ln{(z)}+ {1\over 2}\,\sum_{k=1}^d u_k (z^k-z^{-k})  
\end{array}\right.
}\eeq
The same spectral curve was obtained by Marshakov and Nekrasov in \cite{nekrasov3}, it was derived from finding the optimal profile function which maximizes the free energy $F_0$. It turns out that the extremization equation gives also \eq{loopeqWxab}.

\subsubsection{Consistency of the chiral ansatz}\label{secconsistchans}

We have to check that the $F_g$'s are independent of $N$. The resolvent $W(x)$ is parametrically given by the spectral curve:
\beq
\left\{\begin{array}{l}
\displaystyle x(z) = {N-{1\over 2}\over \sqrt{q}}+\ee{-u_0}\,(z+{1\over z}-u_1) \cr \cr
W(x(z)) =  \sum_{i={\overline{n}}+1}^{N} {1\over \sqrt{q} x(z)-N+i}\,\,\, + 
 \sum_{k=1}^d u_k z^{-k}  + 2\ln{(1+1/z)}  +{1\over 2\sqrt{q} x(z)} \cr
 \qquad \qquad \qquad \qquad  - \sum_{n=1}^\infty {B_{2n}\over 2n (\sqrt{q} x(z))^{2n}}
\end{array}\right.
\eeq
It was proved in \cite{EOFg, EOsym}, that the $F_g$'s are invariant under transformations of the spectral curve which preserve the symplectic form $dx\wedge dy$ (this is why they are called symplectic invariants).
In particular they are invariant in we add a constant to the function $x(z)$, and if we add to $W(x(z))$, any rational function of $x(z)$.
Therefore, the previous spectral curve is symplectically equivalent to:
\beq
\left\{\begin{array}{l}
\displaystyle x(z) = \ee{-u_0}\,(z+{1\over z}-u_1) \cr \cr
y(z) = \ln{(z)}+ {1\over 2}\,\sum_{k=1}^d u_k (z^k-z^{-k})  
\end{array}\right.
\eeq
and this last spectral curve is clearly independent of $N$ and $q$.

Therefore  the $F_g$'s are indeed  independent of $N$, which proves that the chiral ansatz was consistent.

As a bonus, we also find that the $F_g$'s are independent of $q$.

\subsection{Summary: asymptotic expansion}

Using the known properties of matrix models, the logarithm of the following partition function
\beq
Z_N(q,t_k) = \sum_{n(\l)\leq N}\, {\cal P}(\l)\,\, q^{|\l|}\,\, \ee{-\sum_{k\geq 2} {t_k\, q^{1-k\over 2}\over k}\, \, C_k(\l)}
\eeq
can be written as a large $q$ expansion:
\beq
\encadremath{
\ln{Z_N(q,t_k)} \sim \sum_{n=0}^\infty q^{1-n}\, F_n({\cal E}(t_k))
}\eeq
where the $F_n$'s are the symplectic invariants of \cite{EOFg} computed for the spectral curve ${\cal E}(t_k)$ of \eq{spcurvenoq}.
The specral curve, which depends only on the $t_k$'s, is found by solving the transcendental equations \eq{equ0u1gen}.

\medskip

Some explicit examples are treated below in section \ref{secexamplesnq}.

\subsection{Asymptotic shape of large partitions}

\subsubsection{Leading order shape of large partitions}

If we write $z=\ee{i\phi}$ we have the equilibrium density in trigonometric form:
\bea
\rho_{\rm eq}(x)
&=& "\mathop{{\rm lim}}_{q\to\infty}"\,\,\, {1\over \sqrt{q}}\sum_i \left<\delta(x-{h_i\over\sqrt{q}})\right> 
= {y\over i\pi} = {1\over \pi}\left( \phi + \sum_{k=1}^d u_k\,\sin{(k\phi)} \right) \cr
&& \cr
x &=& b+ 2\gamma\,(1+\cos\phi)
\eea
and the integrated density $I=\sqrt{q} \int^{b+4\gamma}_x \rho_{\rm eq} dx$ is:
\bea\label{typicalI}
I &=& {\sqrt{q} \gamma\over \pi}\Big( 2(\sin\phi-\phi \cos\phi) + u_1(\phi-\sin\phi\cos\phi) \cr
&& \qquad \qquad - \sum_{k=2}^d u_k \left({\sin{(k+1)\phi}\over k+1} -{\sin{(k-1)\phi}\over k-1}\right)  \Big) 
\eea
so that $I(\phi=0)=0$ and $I(\phi=\pi)=\overline{n}-{1\over 2}$.

$I(x)$ is the integral of the density, i.e. the number of $h_i$'s between $x$ and $b+4\gamma$. Remember that $h_i=\sqrt{q} x_i$, and thus, $h_{I(x)} = \sqrt{q} x = \sqrt{q} (b+2\gamma(1+\cos\phi))$.
In other words, the inverse function $\sqrt{q}\, x(I)$ gives the typical partition $h_{I(x)}$.
We have parametrically
\beq
\begin{array}{l}
h_I = N-\overline{n}+2\gamma\sqrt{q}(1+\cos\phi) \cr
\cr
I = {\sqrt{q} \gamma\over \pi}\left( 2(\sin\phi-\phi \cos\phi) + u_1(\phi-\sin\phi\cos\phi) - \sum_{k=2}^d u_k \left({\sin{(k+1)\phi}\over k+1} -{\sin{(k-1)\phi}\over k-1}\right)  \right) 
\end{array}
\eeq
The typical $\l_I=h_I+I-N$'s are given by:
\beq
\l_I = I-\overline{n}+2\gamma\sqrt{q}(1+\cos\phi)
\eeq
If we plot $\l_I+I$ as a function of $\l_I-I$, we get the typical shape of the $\pi/4$ rotated partition.
Examples are plotted below in section \ref{secexamplesnq}.

\subsubsection{Subleading fluctuations}

The correlation functions $W_n^{(g)}$'s defined in \cite{EOFg} give all the corrections to the densities and density correlations of the $h_i$'s, to all orders in $\sqrt{q}$.
For instance we have:
\beq
<\sum_i {1\over \sqrt{q}x-h_i}> \sim \sum_{g=0}^\infty q^{-g}\,\, W_1^{(g)}(x)
\eeq
and more generally the cumulant functions expand as:
\beq
\left< \prod_{k=1}^n \left(\sum_i {1\over x_k-{h_i\over \sqrt{q}} }\right) \right>_c \sim  \sum_{g=0}^\infty q^{1-g-n/2}\,\, W_n^{(g)}(x_1,\dots,x_n)
\eeq
where $W_n^{(g)}$'s are the correlators defined in \cite{EOFg} for the spectral curve \eq{spcurvenoq}.

By taking the discontinuity (or by substracting the principal part), we also obtain the correlations of densities $\rho(x)=\sum_i \delta(x-h_i)$:
\beq
<\rho(x)> \sim \sqrt{q}\,{y(x)\over i\pi}  - {1\over 2i\pi}\, \sum_{g=1}^\infty q^{{1\over 2}-g}\,\, W_1^{(g)}(x) 
\eeq
and more generally
\beq
\left<\rho(x_1)\dots \rho(x_n)\right>_c \sim   {i^n \over (2\sqrt{2}\,\pi)^n}\, \sum_{g=1}^\infty q^{1-g-{n\over 2}}\,\, W_n^{(g)}(x_1,\dots,x_n) 
\eeq
The method for computing the $W_n^{(g)}$'s is explained in \cite{EOFg}, and it can easily be implemented on a computer.

\subsection{Examples}\label{secexamplesnq}

\subsubsection{No Casimirs}

We choose all $t_k=0$.

Consider the partition generating function:
\beq
Z_N(q) =  \sum_{n(\l)\leq N} {\cal P}(\l)\,\, q^{|\l|} = 1 + q +{q^2\over 2}+{q^3\over 6}+\dots
\eeq
Equations \eq{equ0u1gen} reduce to $u_k=0$ for all $k$, and $\gamma=1$, and $\overline{n} -{1\over 2} = 2\sqrt{q}$.
The spectral curve is thus:
\beq\label{spcurveqnot}
{\cal E} =
\left\{\begin{array}{l}
\displaystyle x(z) = {N-{1\over 2}-2\sqrt{q}\over \sqrt{q}}+z+{1\over z}+2 \cr \cr
y(z) = \ln{(z)}
\end{array}\right.
\eeq
and therefore:
\beq
\ln{Z_N(q)} \sim \sum_{g=0}^\infty\,\, q^{1-g}\,\, F_g({\cal E})
\eeq
In fact, it is well known \cite{Okounkov1} that when $N\to\infty$, the generating function of partitions is:
\beq
Z_\infty =  \sum_{\l} {\cal P}(\l)\,\, q^{|\l|} = \ee{q}
\eeq
i.e.
\beq
\ln{Z_\infty} = q
\eeq
which shows that:
\beq
F_g({\cal E})=\delta_{g,0}
\eeq

\medskip
\noindent $\bullet$ {\bf Density and typical shape}

We have:
\beq
I = {2\sqrt{q}\over \pi}\,(\sin\phi - \phi \cos\phi)
\virg
\l=I -{1\over 2} + 2\sqrt{q}\cos\phi
\eeq
i.e. by writing $\l-I = 2\sqrt{q} u$ we recover the famous \cite{BDJ, vershik} asymptotic shape of large partitions:
\beq
\l-I = 2\sqrt{q} u
\virg
\l+I = 4\sqrt{q}\, (\sqrt{1-u^2} + u {\rm Arcsin}(u))
\eeq 
$$
{\mbox{\epsfxsize=8.truecm\epsfbox{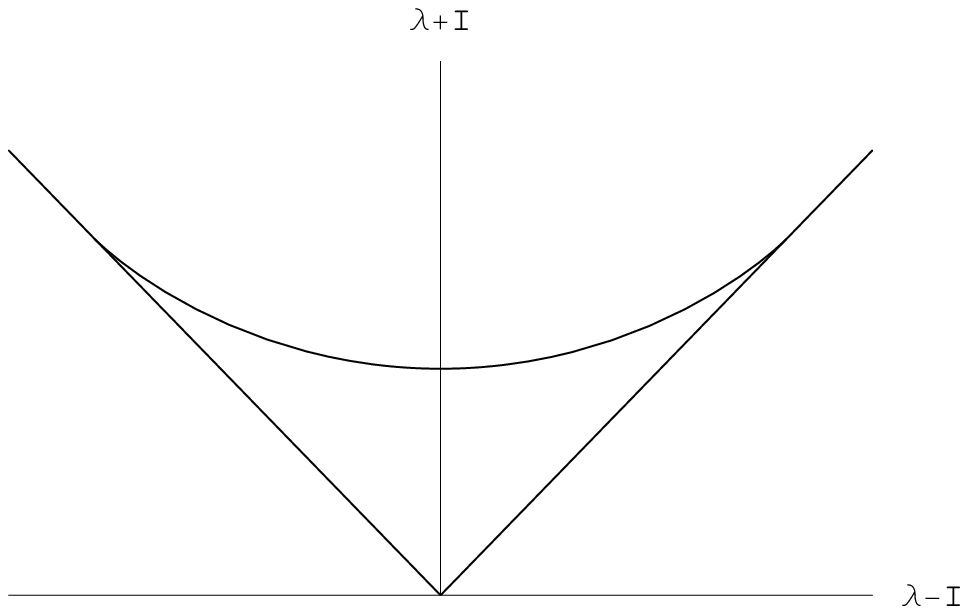}}}\quad
$$

\subsubsection{2nd Casimir}\label{secexCasimir2}

In this example, we choose only $t_2\neq 0$.
Consider the following sum:
\beq
Z_N(q,t_2) =  \sum_{n(\l)\leq N} {\cal P}(\l)\,\, q^{|\l|}\,\,\ee{-{t_2\over 2\sqrt{q}}C_2(\l)}
\eeq
Equations \eq{equ0u1gen} give:
\beq
2u_0 = - t_{2}\,\ee{- u_0}\,  u_1 \virg
u_1 =  t_{2}\,\ee{- u_0}
\eeq
i.e. $u_0$ is determined by:
\beq
-2u_0 \,\, \ee{2u_0}= t_2^2 
\qquad \quad \rightarrow \quad
-2u_0 = \sum_{k=1}^\infty {k^{k-1}\over k!}\,\, t_2^{2k} = t_2^2 + t_2^4 + {3\over 2} t_2^6 + \dots
\eeq
Then we have:
\beq
u_1=\sqrt{-2u_0}
\virg
\gamma=\ee{-u_0}
\virg
\overline{n}-{1\over 2} = (2+u_1)\,\ee{-u_0}\,\, \sqrt{q}
\eeq

The spectral curve is:
\beq
{\cal E}(t_2) = 
\left\{\begin{array}{l}
x(z) =  \ee{-u_0}\,(z+z^{-1}-\sqrt{-2u_0} ) \cr \cr
\displaystyle y(z) = \ln{z}+{\sqrt{-2u_0}\over 2}\,\left(z-{1\over z}\right)
\end{array}\right.
\virg 
-2u_0 = t_2^2 \,\,\ee{-2u_0} 
\eeq
The same spectral curve was already found in \cite{nekrasov3} who computed $F_0$.
Here, we also have the full topological expansion:
\beq
\ln{Z_N(q,t_2)} \sim  \sum_{g=0}^\infty q^{1-g}\,\, F_g({\cal E}(t_2))
\eeq
For example we find (using mathematica with the definitions of  \cite{EOFg}):
\bea
F_1 &=& {1\over 24}\ln{\left(\ee{-2u_0}\,(1+2u_0)\right)} \cr
F_2 &=& {\ee{2u_0}\over 180}\, {u_0^3 \,(1-12 u_0)\over (1+2u_0)^5} \cr
& \vdots &
\eea

As an example, we plot the typical shape of a partition with $t_2=0.2$:
$$
{\mbox{\epsfxsize=8.truecm\epsfbox{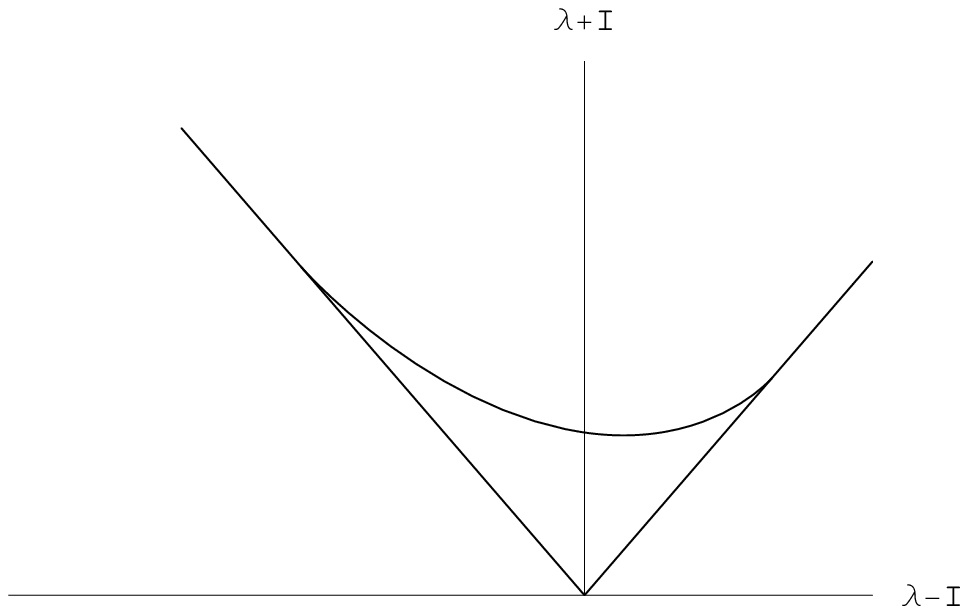}}}\quad
$$
and corrections to the limit shape are given by:
\bea
\left< \sum_i {1\over \sqrt{q} x(z) - h_i}\right>
&=& \om(z)  + {1\over q}\,\om_1^{(1)}(z) + O(q^{-2})
\eea
where
\bea
\om_1^{(1)}(z)
&=& {1\over 24\,e^{-u_0}\,{\left( 1 + 2 u_0 \right) }^2\,{\left( z^2 -1  \right) }^4}\,\,\,
\Big( (1+z^2)(1-14 z^2+z^4) 
- 24\,\,{(-2u_0)}^{\frac{3}{2}}\,z^3  \cr
&& \qquad  +     2\,{\sqrt{{-2u_0}}}\,z\,( 1 + 10\,z^2 + z^4 )  + 
    4\,u_0\,(1+z^2)(1-8 z^2+z^4) \Big)
\eea

\subsection{Phase transitions and number of cuts}
\label{secphasetr}

In section \ref{secchiralansatz}, we made the assumption that $\td{W}(x)$ has only one cut $[0,a]$.
Then we could compute the density of the $h_i$'s on the cut.
It may happen that the density we compute this way is not positive, which means that our 1-cut assumption was possibly wrong. 

For example, in the example of section \ref{secexCasimir2}, a "pure gravity" phase transition (i.e. conformal field theory (3,2), characterized by the Painlev\'e I equation)  occurs when $t_2=\ee{-1}$ (i.e. $u_0=-{1\over 2}$).

In fact, all situations may occur as we vary the parameters $t_k$'s: cuts may split, merge, shrink, or appear, they can move to the complex plane, and branch so that they form trees. Most of those transitions have been described in the random matrix theory literature, and take place also in this particular matrix model.

In principle, the correct assumption, is the one which minimizes $\Re F_0$.

It would be interesting to study the phase transitions between different regimes of the $t_k$'s, and see how the cuts change, and thus how the shape of typical large partitions change. This would be merely an adaptation of mostly known phase transitions of matrix models, specialized to this case.
We won't do it in this article.

\bigskip

However, a non-positive density is not a problem when the partition functions we consider are only formal series.

The importance of the 1-cut assumption, comes from the fact, that most often, this is the assumption which gives the formal generating series which enumerates some objects.
Formal generating series are very often obtained from Feynman's method, i.e. a perturbative expansion near a minimum of an action. 
We keep the quadratic part of the action in the exponential, and expand the non-quadratic terms, so that we end up having to compute series of polynomial moments of Gaussian integrals, which are represented diagrammatically through Wick's theorem. This method naturally associates combinatorics objects to integrals \cite{eynform}.

A 1-cut case means that all eigenvalues of the random matrix are expanded near the same minimum, whereas multicut cases mean that we expand near several minima.
Multicut cases lead to formal series which enumerate objects of different types (one type for each minimum, i.e. for each cut). This is less natural than 1-cut cases where we enumerate only objects of the same type.

Thus, for formal series, the correct assumption on the number of cuts, is not obtained by minimizing $\Re F_0$, but is obtained from the type of objects which we wish to enumerate, i.e. near which minimum we want to expand.

\bigskip

Here in this article, we focused on the 1-cut case for simplicity, but it is clear that the entire method would work for any number of cuts. The symplectic invariants of \cite{EOFg} were indeed defined for any number of cuts.

\subsubsection{Universal regimes}

Since we have a matrix model, we find the same universal regimes as usual in matrix models.
Correlation functions in various regimes are obtained from some universal kernels.
Also, it was found in \cite {EOFg} (cf theorem 8.1 in \cite{EOFg}) how to study the vicinity of phase transitions of the $F_g$'s.
 
For instance it is clear that we have "sine-kernel" in the bulk of the spectrum at microscopic scaling, "Szeg\"o kernel" in the bulk at macroscopic scaling, "Airy kernel" and Tracy-Widom law \cite{TW} (which is also the conformal minimal theory $(1,2)$) near regular turning points, and we have the Kadamtsev-Petviashvili (KP) kernels (i.e. minimal conformal field theory $(p,q)$) near singular points of the type $y\sim x^{p/q}$, and polynomial kernels of \cite{eynbirthcut} when a new cut opens.
In general, the KP kernel for a $(p,q)$ critical point, is the kernel (defined in section 9.1. in \cite{EOFg}) of the limit spectral curve (see section 10.3 of \cite{EOFg}), of equation
\beq
{\cal E}_{p,q} = \left\{\begin{array}{l}
x(z) = Q(z)\virg \deg Q=q \cr
y(z) = P(z)\virg \deg P=p \cr
\end{array}\right.
\eeq 

However, we shall not study the details of those universal regimes and transitions in this article.

\vspace{1cm}

\section{The q-deformed Plancherel measure}

In this section, we repeat the same transformation into a matrix integral, chiral ansatz and topological expansion, for the $q$-deformed Plancherel measure.

\medskip
We compute:
\beq
Z_N(q,t_k) = \sum_{n(\l)\leq N}\,{\cal P}_q(\l)\,\,\,\, \ee{-\,{1\over g_s}\,\sum_{k=1}^{d+1}\, {t_k\, g_s^{k}\over k}C_k(\l)}
\virg
q=\ee{-g_s}
\eeq
in the small $g_s$ topological expansion regime:
\beq
\ln{Z_N(\ee{-g_s},t_k)} \sim \sum_{g=0}^\infty g_s^{2g-2}\,\, F_g
\eeq

We shall discuss an important application of this model to topological string theory in section \ref{secXp} below.

\subsection{Transformation into a matrix integral}

Let
\beq
q = \ee{-g_s} 
 \virg  T = g_s (N-{1\over 2})
\eeq
\medskip

Let the contour ${\cal C}_q$ be a circle of radius $r$ with  $|q^{-1}|>r>1$ centered in $0$.
In particular it encloses all the $q^h$ with $h\in {\mathbb N}$.
$$
{\mbox{\epsfxsize=4.truecm\epsfbox{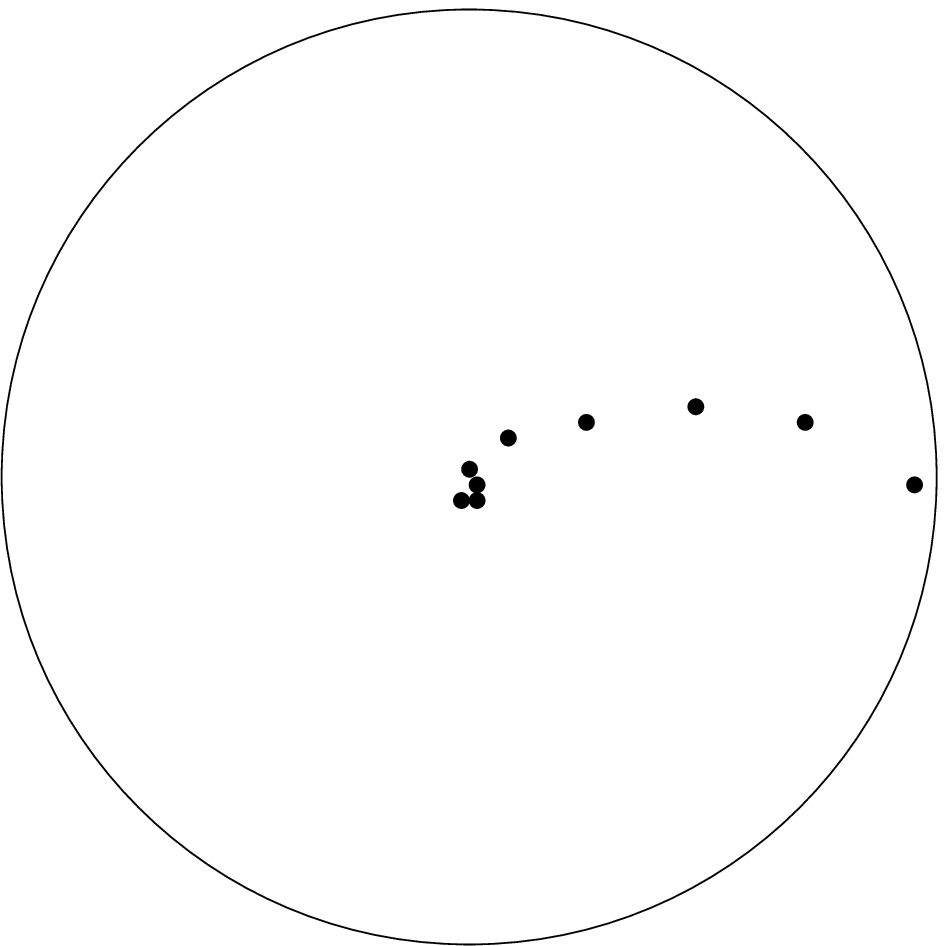}}}
$$

Consider the $q$-series:
\beq
g(x) = \prod_{n=1}^\infty (1-{1\over x}q^n)
\eeq
It vanishes when $x=q^h$ with $h\in {\mathbb N}^*$, and we have:
\beq
g(q^h)=0
\virg
g'(q^h) = - {g(1)^2\, \ee{i\pi h}\, q^{-h(h-1)/2} \over q^h(1-q^h)\, g(q^{-h})} 
\eeq
and thus the function:
\beq
f(x) =  - {g(1)^2\, \ee{-{i\pi\over g_s}\,\ln{(x)} }  \ee{{1\over 2g_s}\,(\ln{x})^2} \over (1-x)\, \sqrt{x}\, g(x) g(1/x)}
\eeq
has simple poles with residue $1$ when $x=q^h$ for $h\in {\mathbb N}$ (one easily sees that it has a pole of residue 1 at $x=1$ as well).

Notice that if $x=q^h$ with $h\in {\mathbb{N}}$, we have:
\beq
[h]! = \prod_{j=1}^h (q^{-j/2}-q^{j/2}) = q^{-h(h+1)/4}\,\,  {g(1)\over g(1/x)}
\eeq

Therefore we have:
\bea
&& \oint_{{\cal C}^N}\, dx_1\dots dx_N\,\, \prod_{i<j} (x_i-x_j)^2\,\,\prod_{i} f(x_i)\,{\ee{-A(\ln{(x_i)}/\ln{q})}\,\, {g(1/x_i)^2\over g(1)^2}\,\, {\ee{{\ln{(x_i)}\ln{(qx_i)}\over 2\ln{q}}}} \over x_i^{N-1}}  \cr
&=& \sum_{h_1,\dots,h_N=0}^\infty \,\, \prod_{i<j} (q^{h_i}-q^{h_j})^2\,\,\prod_{i} {\ee{-A(h_i)}\over [h_i]!^2} \,\, q^{(1-N)h_i}  \cr
&=& N!\,\sum_{h_1>\dots>h_N\geq 0} \,\, {\prod_{i<j} (q^{h_i}-q^{h_j})^2\over [h_i]!^2}\,\,\prod_{i} \ee{-A(h_i)} \,\,q^{(1-N)h_i}  \cr
&=& N!\,\sum_{h_1>\dots>h_N\geq 0} \,\, {\prod_{i<j} (q^{h_i-h_j\over 2}-q^{h_j-h_i\over 2})^2\over [h_i]!^2}\,\,\prod_{i} \ee{-A(h_i)}   \cr
\eea

In other words, the $q$-deformed Plancherel measure partition function is a matrix integral:
\beq
\sum_{h_1>\dots>h_N\geq 0} \,\, {\prod_{i<j} [h_i-h_j]^2\over [h_i]!^2}\,\,\prod_{i} \ee{-A(h_i)}   
= {1\over N!}\,\int_{H_N({\cal C}_q)}\, dM\, \ee{-{1\over g_s} \Tr V(M)}
\eeq
with the potential:
\bea
{1\over g_s} V(x) 
&=& \ln{g(x)}-\ln{g(1/x)} +(N-1)\ln{x}+{i\pi\over g_s}\,\ln{x} + \ln{(1-x)} \cr
&& +{1\over g_s}\sum_k {(-1)^k\,t_k\over k}\,\, (T+\ln{x})^k 
\eea

\medskip
{\noindent \bf Remark:}
since the contour ${\cal C}_q$ is a circle of radius $r$, The ensemble $H_N({\cal C}_q)$ is (up to a scaling $r$) the ensemble of unitary matrices:
\beq
H_N({\cal C}_q) = r\, U(N)
\eeq
It was already noticed \cite{baik, johansson}, that sums over partitions with the Plancherel measure have many common properties with unitary matrix models, and here the connection becomes very apparent.

\bigskip

\subsection{The potential}

We use the following asymptotic expansion of the $q$-series (for $|q^{-1}|> |x|\geq 1$):
\beq
\ln{(g(x))} = -{1\over g_s} \sum_{m=0}^\infty  Li_{2-m}(x^{-1}) \,  {B_m\over m!} \,\, g_s^m 
\eeq
where $B_m$ is the $m^{\rm th}$ Bernouilli number ($B_0=1, B_1=-{1\over 2}, \dots$, and $B_{2m+1}=0$ if $m\geq 1$), and $Li_n(x)=\sum_{k=1}^\infty x^k/k^n$ is the polylogarithm function.
We have:
\beq
x{g'(x)\over g(x)} = {1\over g_s} \sum_{m=0}^\infty  Li_{1-m}(x^{-1}) \,  {B_m\over m!} \,\, g_s^m 
\eeq
The polylogarithm functions have many properties, in particular:
\beq
Li'_n = {1\over x}Li_{n-1}
\eeq
\beq
Li_{-m}(1/x) = (-1)^m Li_{-m}(x) \qquad m>0
\eeq
\beq
Li_{1}(x) =-\ln{(1-x)}
\virg
Li_0(x)={x\over 1-x}
\eeq
and $Li_{-m}$ with $m>0$ is a rational fraction with poles at $x=1$ and which vanishes at $x=0$ and at $x=\infty$.

Therefore:
\bea
x V'(x) 
&=&   \sum_{m=0}^\infty  (Li_{1-m}(x)+Li_{1-m}(x^{-1})) \,  {B_m\over m!} \,\, g_s^m  \cr
&& +(N-1)g_s+ i\pi + g_s {x\over x-1} - \,\sum_{k=0}^d (-1)^k\, t_{k+1} (T+\ln{x})^k \cr
&=&   (-2\ln{(1-x)} + \ln{x} - i\pi)  + {g_s\over 2}      \cr
&& + 2 \sum_{m=1}^\infty  Li_{1-2m}(x) \,  {B_{2m}\over 2m!} \,\, g_s^{2m}  \cr
&& + (N-1)g_s+  i\pi +  g_s {x\over x-1} - \,\sum_{k=0}^d (-1)^k\, t_{k+1} (T+\ln{x})^k \cr
&=&   -2\ln{(1-x)} + \ln{x}+T  - \,\sum_{k=0}^d (-1)^k\, t_{k+1} (T+\ln{x})^k    \cr
&& + {g_s x \over x-1} + 2 \sum_{m=1}^\infty  Li_{1-2m}(x) \,  {B_{2m}\over 2m!} \,\, g_s^{2m}   \cr
\eea
Notice that terms on the last line are rational fractions with poles at $x=1$, and which vanish at $x=0$ and at $x=\infty$.

\subsection{Restriction to 2 Casimirs}

In principle, the recipe above, should allow to find the spectral curve for any $t_k$'s, but for simplicity,
from now on, we shall assume that $t_k=0$ if $k\geq 3$, and we write:
\beq
t_1=t \virg t_2=p-1
\virg t_3=t_4=t_5=\dots=0
\eeq
The potential is then:
\bea
x V'(x) 
&=&   -2\ln{(1-x)} + p (\ln{x}+T)  -t     \cr
&& + 2 \sum_{m=1}^\infty  Li_{1-2m}(x) \,  {B_{2m}\over 2m!} \,\, g_s^{2m} + {g_s x \over x-1}  \cr
\eea
Those notations $(t,p-1)=(t_1,t_2)$ are motivated by the application to algebraic geometry \cite{CGMPS, Marcosconj}, namely the computation of the Gromov-Witten invariants of $X_p$, which we present in section \ref{secXp}, see eq.\ref{ZXp}.

\smallskip
Also, a motivation for the assumption $t_k=0$  for $k\geq 3$, is that the spectral curve is much simpler in that case. For higher $t_k$'s, we would need to play with tedious identities among polylogarithms, and thanks to this assumption, we don't need higher polylogarithms than $Li_1(x)=-\ln{(1-x)}$.

\subsection{Chiral ansatz and dependence on $N$}

We shall not repeat the discussion of sections \ref{secarcticcircle}, \ref{secchiralansatz} and \ref{secconsistchans}.
We let it to the reader (following the same steps as in the undeformed Plancherel measure) to check that the spectral curve has an arctic circle property, i.e. all $h_i$'s with $i>\overline{n}$ are frozen, and we can choose a chiral ansatz \cite{chiral, CGMPS} such that the $F_g$'s don't depend on $N$.
Therefore we chose $N=\overline{n}$, such that the edge of the small $q$ average distributions of the $h_i$'s, is at $h=0$, i.e. $x=1$.
The authors of \cite{CGMPS} checked the consistency of the chiral ansatz, by verifying that $F_0$ is independent of $T$ (there is a $T^2$ scaling between ours and that of \cite{CGMPS}).

In other words, in an appropriate regime of $t$ and $p$, we may choose $T$, such that the spectral curve corresponds to a 1-cut distribution, with edge at $x=1$.

\subsection{Spectral curve}

Like in section \ref{spcurverecipe}, the spectral curve is determined by:
\beq\label{loopeqWVq}
W(x+i0)+W(x-i0) = V'(x)
\virg \forall\,\, x\in [a,1]
\eeq
and $W(x)$ is analytical outside the cut $[a,1]$, and at large $x$:
\beq
W(x) \sim {\overline{n} g_s\over x}
\eeq

\bigskip

Since we look for a 1-cut solution, we again use the Zhukovski parametrization and write:
\beq
x(z) = 1 -  \gamma\,(z+2+{1\over z}) \virg a = 1-4\gamma
\eeq
Notice that each value of $x$ corresponds to two values of $z$, namely $z$ and $1/z$.
We call the domain $|z|>1$ the physical sheet, and $|z|<1$ the second sheet.

Define $z_0$ such that $|z_0|\geq 1$ and $x(z_0)=0$, that is:
\beq
{1\over \gamma} = (1+z_0)(1+1/z_0) 
\eeq

\medskip

Then, the resolvent is $W(x(z))=\om(z)$:
\bea
x(z) \om(z)
&=&  -2\, (\ln{(1+1/z)}-\ln{(1+1/z_0)}) \cr
&& + p\,(\ln{(1-1/z z_0)}-\ln{(1-1/z_0^2)})  \cr
&& +  \sum_{m=1}^\infty  Li_{1-2m}(x(z)) \,  {B_{2m}\over 2m!} \,\, g_s^{2m} + {g_s x(z)\over 2(x(z)-1)}  \cr
\eea
Indeed, one easily checks that $\om(z)$ has no singularity in the physical sheet $|z|>1$.
The condition $\om(z) \sim {g_s \overline{n} \over x(z)}$ at large $z$ implies:
\beq
2\ln{(1+1/z_0)} - p\ln{(1-1/z_0^2)}  = T = g_s (\overline{n}-{1\over 2})
\eeq
Then, the loop equation \eq{loopeqWVq} implies:
\bea
x(z)(\om(z)+\om(1/z) -V'(x(z))) &=&0 \cr
&=&  2\, (\ln\gamma+2\ln{(1+1/z_0)}) \cr
&& -p\,(\ln{(\gamma z_0)}+2\ln{(1-1/z_0^2)}) \cr
&&  +t-pT \cr
\eea
therefore $z_0$, $T$ and $\gamma$ are determined by the 3 equations:
\beq
\left\{\begin{array}{l}
0 =   2\, (\ln\gamma+2\ln{(1+1/z_0)})  -p\,(\ln{(\gamma z_0)}+2\ln{(1-1/z_0^2)})   +t-pT \cr
T= 2\ln{(1+1/z_0)} - p\ln{(1-1/z_0^2)}   \cr
{1\over \gamma} = (1+z_0)(1+1/z_0) 
\end{array}\right.
\eeq
%
and after a little bit of algebraic manipulations,
$z_0$ is determined as a function of $t$ and $p$ by:
\beq\label{eqtz0}
\encadremath{
\ee{-t} = {1\over z_0^2}\,\,(1-{1\over z_0^2})^{p(p-2)}
}\eeq
One should notice that this equation is symmetric in $p\leftrightarrow 2-p$, which reflects the symmetry of the Calabi-Yau space $X_p \leftrightarrow X_{2-p}$.
The same equation was found in \cite{CGMPS} where the authors wrote $w=1-{1\over z_0^2}$ (see equation 4.34 in \cite{CGMPS} or equation 3.14 of \cite{Marcosconj}).

Order by order in $\ee{-t}$ we have (using Lagrange inversion formula):
\bea
{1\over z_0^2} 
&=& \ee{-t}+\sum_{k=2}^\infty {\ee{-kt}\over k!} \,\prod_{j=0}^{k-2}\, (kp(p-2)+j) \cr
&=& \ee{-t} + p(p-2) \ee{-2t} + {p(p-2)(3p(p-2)+1)\over 2}\,\ee{-3t} +  \dots
\eea
We also determine $T$ and $\gamma$ through:
\beq
\ee{-T} ={(1-{1\over z_0})^p\over (1+{1\over z_0})^{2-p}} 
\virg
{1\over \gamma} = (1+z_0)(1+1/z_0) 
\eeq
The arctic circle is at $N=\overline{n} = {1\over 2}+{T\over g_s}$.

Finally, our spectral curve is:
\beq\label{spcurveq}
\encadremath{
\curve(t,p) = 
\left\{
\begin{array}{l}
x(z)  =  {(1-{z\over z_0})(1-{1\over z z_0})\over (1+{1\over z_0})^2} \cr
\cr
y(z) = {1\over x(z)}\,\left(- \ln{z}  + {p\over 2}\,\ln{\left(1-z/z_0\over 1-1/z z_0\right)}  \right) 
\end{array}\right.
\quad ,\,\, \ee{-t} = {1\over z_0^2}\,\,(1-{1\over z_0^2})^{p(p-2)}
}\eeq

One may check that this spectral curve is identical to that of \cite{CGMPS, Marcosconj}
(see eq. 3.31 in \cite{Marcosconj}), therefore it gives the same $F_g$'s.

This proves the conjecture of \cite{Marcosconj}.


\subsection{Topological expansion}

\subsubsection{Topological expansion}

Since $Z$ is by definition a formal power series in $g_s$ (it is the generating function of Gromov-Witten invariants of $X_p$, see section \ref{secXp}), and is given by a matrix model, using \cite{EOFg}, we find that $Z$ has a topological expansion
\beq
\ln{Z} = \sum_{g=0}^\infty g_s^{2g-2}\, F_g(\curve)
\eeq
where the coefficients $F_g(\curve)$ are the symplectic invariants defined in \cite{EOFg}.

\medskip
Let us compute the few first orders (the first orders were computed in \cite{CGMPS}):

$\bullet$ $F_0$ is found from the general theory of \cite{EOFg} through its 3rd derivative:
\beq
{\partial^2 F_0\over \partial t^2} = -\ln{(1-{1\over z_0^2})} = \sum_k {\Gamma(k(p-1)^2)\over k!\, \Gamma(kp(p-2)+1)}\,\ee{-tk}
\eeq

It can be checked that this leads to the expression found in \cite{CGMPS, Marcosconj}:
\beq
F_0 = p(p-2)Li_3({1\over 1-z_0^2}) + (p-1)^2 Li_3(1/z_0^2) - {p(p-2)(p-1)^2\over 6}\, (\log{(1-{1\over z_0^2})})^3
\eeq
For instance if $p=0$ or $p=2$ we have $F_0''' = \sum_k \ee{kt}$, i.e. $F_0 = \sum_k {\ee{kt}\over k^3}=Li_3(\ee{t})$.
And for $p=1$ we have $F_0''' = \sum_k (-1)^k\, \ee{kt}$, i.e. $F_0 = Li_3(-\ee{t})$.

$\bullet$ $F_1$ is equal to (according to \cite{EOFg}):
\beq
F_1= {1\over 24}\ln{\left(\gamma^2 y'(1)y'(-1)\right)} 
= {1\over 24}\ln{\left(z_0^2((p-1)^2-z_0^2)\over (1-z_0^2)^3\right)}
\eeq
which is the same as in \cite{Marcosconj}.

$\bullet$ $F_2$ is obtained from \cite{EOFg} (using a few minutes basic Mathematica programm):
\bea
F_2 &=&
{1\over        2880\,{\left( z_0^2- (p-1)^2  \right) }^5}\,\,\Big(
(p-1)^8 (-1+12 z_0^2-12 z_0^4)  \cr
&& + (p-1)^6 z_0^2 (-5+z_0^2+2z_0^4)  + 35 (p-1)^4 z_0^4 (-1+z_0^2) \cr
&& + (p-1)^2 z_0^4 ( 2+z_0^2-5z_0^4) + z_0^6(12-12 z_0^2 + z_0^4)
\Big)\cr
\eea

$\bullet$ Higher $F_g$'s can be obtained easily with a higher computing time...

\subsubsection{Comparison with the conjecture of \cite{Marcosconj}}

The authors of \cite{CGMPS} have computed the spectral curve as the solution of the large $N$ saddle point equation for the $h_i$'s. Then, Marcos Mari\~ no \cite{Marcosconj} computed the first few orders $F_0$ and $F_1$ for that spectral curve, and conjectured that the Gromov-Witten invariants were generated by the symplectic invariants $F_g$ of the same spectral curve for all $g$.
However, our approach is rather different, and it is worth to comment on "why it works ?".
It is due to a few miracles.

In fact, in their computation, the authors of \cite{CGMPS}, worked in the leading large $N$ limit, and they made several approximations.

- The first of them, is that they set $T=Ng_s$ instead of $T=(N-{1\over 2})g_s$. This makes no difference to leading order, but it can make a big difference to subleading orders.

- The second of them, is that they neglected all terms with a $1/N$ behaviour in the Plancherel measure, keeping only $Li_2$ and $Li_1$, they neglected all the Bernouilli series.

- The third of them, is that they neglected the discreteness of the $h_i$'s, and replaced them by continuous variables, i.e. they replaced sums by integrals, whereas here in this article, we encoded the discreteness of the $h_i$'s by the $q-$series $g(q^h)$, which forces (without approximation) the $h_i$'s to be integers. A miracle is that the $g$-function combines very well with the Plancherel measure.
\smallskip

Out of those approximations, they deduced a spectral curve, and quite miraculously, it turns out that it is indeed the same curve as we found here, modulo symplectomorphisms. 
In some sense the 3 approximations compensate each other, and they indeed found the right spectral curve which gives the answer to all orders.
This miracle, is of course due to the very special nature of Plancherel measure, it would not have worked with another arbitrary measure.


\subsection{Symplectic invariance and mirror symmetry}

Here, we use the fact that the $F_g$'s are symplectic invariants, i.e. two spectral curves which have the same symplectic form $dx\wedge dy$ have the same $F_g$'s.

\medskip
Notice that the spectral curve \eq{spcurveq} has the form:
\beq
\curve(t,p) = 
\left\{
\begin{array}{l}
x(z) = u(z)  \cr
y(z) = {1\over x(z)}\, \ln{v(z)} 
\end{array}\right.
\eeq
where both $u$ and $v$ are rational fractions of $z$.

The symplectic form $dx\wedge dy$ is thus:
\beq
dx\wedge dy = d\ln{u(z)}\, \wedge \, d\ln{v(z)}
\eeq
therefore the spectral curve $\curve(t,p)$ is symplectomorphic to the following new spectral curve:
\beq
\td\curve(t,p) =
\left\{
\begin{array}{l}
x(z) = \ln{u(z)} \qquad \virg u(z)= \gamma z_0 (1-{z\over z_0})(1-{1\over z z_0})  \cr
y(z) = \ln{v(z)} \qquad \virg v(z)={1\over z}\, \left( 1-{z\over z_0}\over 1-{1\over z z_0} \right)^{p\over 2}
\end{array}\right.
\eeq
which is an algebraic curve in $\ee{x}$ and $\ee{y}$, i.e. there exists a polynomial $H(u,v)$ such that:
\beq
H(\ee{x},\ee{y})=0
\eeq
This is typically an equation of a mirror Calabi-Yau 3-fold in type B topological string theory \cite{horivafa, Agvafa, bookmirror, reBmodel}.

\medskip

Remark: notice that the transformation $(u,v)\to(u^f,v)$ is also a symplectic transformation which leaves the $F_g$'s unchanged. It corresponds to the so-called framing transformation.

\subsection{Mirror curve}

The curve $H(\ee{x},\ee{y})=0$ can be written by eliminating $z$:
\beq
(\ee{y}+\ee{-y})(1+{1\over z_0})^{p/2} \, \ee{xp/2}= \sum_j \pmatrix{p\cr j}\,{(-1)^j\over z_0^j}\,\, T_{j-1}((1+z_0)(1+{1\over z_0})(1-\ee{x})-2)  
\eeq
where $T_j$ is the $j^{\rm th}$ Tchebychev polynomial, such that $T_j(z+z^{-1}) = z^j+z^{-j}$.

\bigskip


$\bullet$ Example $p=0$ ($z_0 = \ee{t/2}$):
\beq
v+{1\over v}+2= (1-u)(1+\ee{t\over 2})(1+\ee{-t\over 2})
\eeq
i.e.
\beq
0= (1+\ee{y})(1+\ee{-y}) + (\ee{x}-1)(1+\ee{t\over 2})(1+\ee{-t\over 2})
\eeq

\bigskip

$\bullet$ Example $p=1$ ($z_0^2 -1 = \ee{t}$):
\bea
v^2+{1\over v^2} 
= (1+z_0)^2 u + (1-z_0)^2 u^{-1} - 2 z_0^2
\eea

$\bullet$ Example $p=2$ ($z_0 = -\ee{t/2}$):
\bea
v+v^{-1}+2 = (1-u^{-1}) (1+\ee{t\over 2})(1+\ee{-t\over 2})
\eea
which is symplectically equivalent to the case $p=0$ by changing $u\to u^{-1}$.

$\bullet$ Example $p=-1$ ($z_0^{-2}(1-z_0^{-2})^3  = \ee{-t}$):
\bea
v^2+v^{-2} = 
 (1+{1\over z_0})^2 u + (1-{1\over z_0})^2 u^{-1} - {2\over  z_0^2}
\eea

\section{Some applications}
\label{secapplis}

In this section we discuss some applications of our method.

\medskip

Important applications of sums of partitions with the Plancherel measure include growth phenomenon (crystals), length of increasing subsequences of a random sequence, Totally asymmetric exclusion processes, and algebraic geometry through the computation of Gromov-Witten invariants and Hurwitz numbers.

Many of those applications are based on standard Young tableaux.

\medskip
A {\bf standard Young tableaux}
, is a Ferrer diagram filled with numbers such that the numbers are increasing along each row and along each column.
The number of standard Young tableaux of shape $\l$, is ${\rm dim}(\l)$.

\subsection{Longest increasing subsequence}

Let $\sigma\in \Perm_k$ be a permutation of $k$ elements. Call $l(\sigma)$ the length of the longest increasing subsequence of $\sigma$.

The Robinson-Schensted algorithm \cite{Robinson, ssuites1} gives a bijection between the group of permutations $S_k$ and the set of pairs of Standard Young Tableaux with $k$ boxes $(T,T')$ with the same shape $\l$ (${\rm shape}(T)={\rm shape}(T')=\l$, $|\l|=k$).
This bijection illustrates the formula:
\beq
k! = \sum_{\l\, , \, |\l|=k}\, ({\rm dim}(\l))^2
\eeq
The length of the longest increasing subsequence is in bijection with the length of the partition $n(\l)$.
The uniform probability law on permutations induces the Plancherel law on partitions.
Therefore, the probability law of the length of the longest increasing subsequence of a random permutation (uniformly chosen), is the probability law of $n(\l)$ with the Plancherel measure \cite{Aldous, baik, BDJ, johansson, kerov1, kerov2, kerov3, Logan, ssuites1, Planchrel1, vershik}.
What one needs to compute is the expectation value:
\beq
{\mathbb E}(l(\sigma)\leq N) = {\mathbb E}_{k,N} = {1\over k!}\sum_{\sigma\in \Perm_k\, , l(\sigma)\leq N}\,\, 1 
= k!\, \sum_{\l, \, |\l|=k,\, n(\l)\leq N}\,\, {\cal P}(\l)
\eeq
Instead of working with permutations of a given number of elements $k=|\l|$, it may be more convenient to work with a "Poissonized"  version (grand canonical ensemble), where we sum over all numbers $k=|\l|$,  with a "chemical potential" of the form $q^{|\l|}$.
In the end, we need to compute:
\beq
Z_N(q) = \ee{-q}\,\sum_k {\mathbb E}_{k,N}\, {q^k\over k!}\,   = \ee{-q}\,\sum_{\l,\,n(\l)\leq N} {\cal P}(\l)\, q^{|\l|}
\eeq

Baik, Deift and Johansson \cite{BDJ} proved that $Z_N(q)$ is related to a Haenkel determinant, and that in the large $q$ limit, the distribution of the length of the longest increasing subsequence converges towards the Tracy-Widom law \cite{TW}.
In particular they could find the "typical shape" of a large partition. For instance they found that the typical size $k=|\l|$ is given by:
\beq
<|\l |> = q
\eeq
and the typical length $l(\sigma)=n(\l)$ is given by:
\beq
<l(\sigma)>=<n(\l)> = 2\sqrt{q}
\eeq
and typical fluctuations are of order $q^{1/ 6}$.
\beq
<l(\sigma)-2\sqrt{q}> \sim O(q^{1/6})
\eeq
Here, in this article, we compute not only the large $q$ leading order of $Z_N(q)$, but the whole $1/q$ expansion of $Z_N(q)$ for large $q$.

\subsection{Growing/melting crystal}

The Robinson-Schensted algorithm also gives a simple growth model for a  2-dimensional crystal.

The crystal is represented by $\pi/4$ the rotated partition $\l$.
At time $t\in {\mathbb N}$, we drop a new box, falling from the sky at a random position $\sigma(t)$. When the box hits the crystal, it slides to the left or to the right until it can no longer move.
If we assign the box, the time $t$ at which it arrives (resp. the position $\sigma(t)$ from which it came), we clearly obtain at each unit of time a standard Young tableau, and the number of ways of obtaining a given standard tableau of shape $\l$ after $|\l|$ units of time, is the number of permutations $\sigma$ which can lead to it, i.e. it is the number of standard Young tableaux of shape $\l$, i.e. it is ${\rm dim}(\l)$.
The 2 standard tableaux we obtain from assigning the box $t$ or $\sigma(t)$ are precisely the 2 tableaux of the Robinson-Schensted bijection.

$${\mbox{\epsfxsize=8.truecm\epsfbox{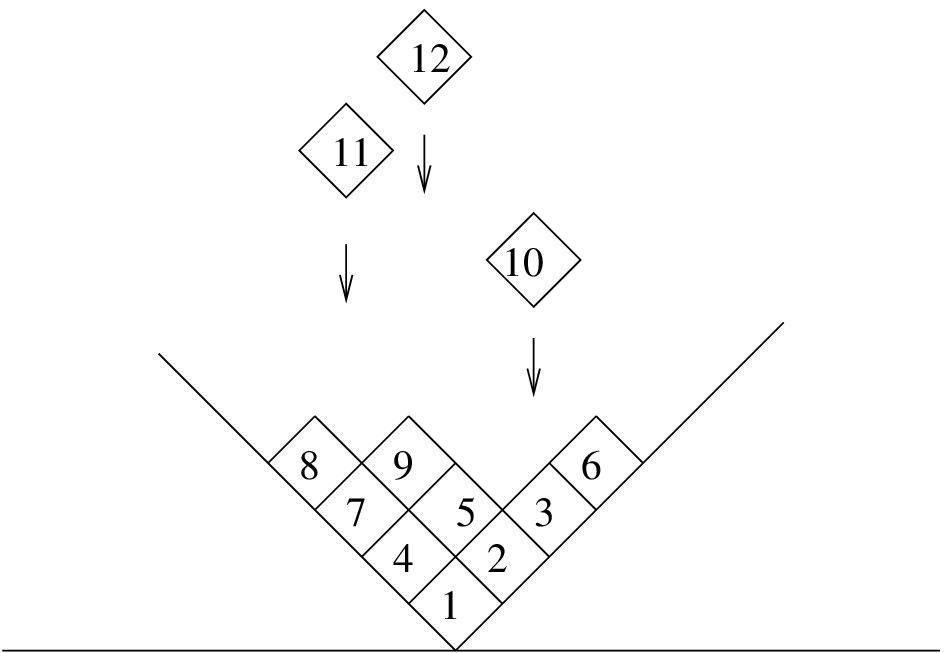}}}$$

This is a model of a growing crystal.

The other way around, it is also a melting crystal, where at each unit of time only boxes which are at corners can evaporate. If we label each box by the time at which it evaporates, we get a standard tableau.

\smallskip
Therefore it is natural to associate to each partition, a weight which is its entropy, namely the number of ways of obtaining it, i.e. ${\rm dim}(\l)^2$. If the number of boxes is not fixed, we may consider a grand-cannonical ensemble, with a weight depending on $\l$, for instance of the form $q^{|\l|}$. 
Again, the partition function we need to compute is of the form:
\beq
Z(q) = \sum_\l\, {\cal P}(\l)\,\, q^{|\l|}
\eeq
and we are also interested in expectation values of various observables, typically expectation values of Casimirs.

\smallskip

This analogy between partitions and crystals has been very useful and fruitful, and has generated a considerable amount of works and discoveries in physics and mathematics \cite{spohn, kerov3, baik, ferrari1, ferrari2, Okounkov1}.
 
\subsection{T.A.S.E.P.}

The acronym T.A.S.E.P. stands for totally asymmetric exclusion process.
It is a famous model of statistical physics, where particles are at integer positions on the real axis \cite{liggett, derrida, derrida2, Mallick}.
At each unit of time, we choose one particle, and with a certain probability, either it stays at the same place, or it jumps to the next position to the right if unoccupied. This model is very important because it is the simplest "out of equilibrium" statistical physics model.

\medskip
There is a bijective mapping between this model and the growing crystal.
Indeed, in the growing crystal, at each time $t$, we have a partition $\l(t)$ of weight $|\l(t)|=t$.
If we consider the $\pi/4$ rotated partition, we have for each $t$, a set of ordered integers $h_1(t)>h_2(t)>\dots>h_N(t)\geq 0$. If we interpret the integer $h_i(t)$ as the positions of the $i^{\rm th}$ particle at time $t$, we see that those particles follow a T.A.S.E.P., whose initial condition at $t=0$ is a "Dirac see" (${\mathbb Z}_-$ full of particles, ${\mathbb N}$ empty).
$${\mbox{\epsfxsize=12.truecm\epsfbox{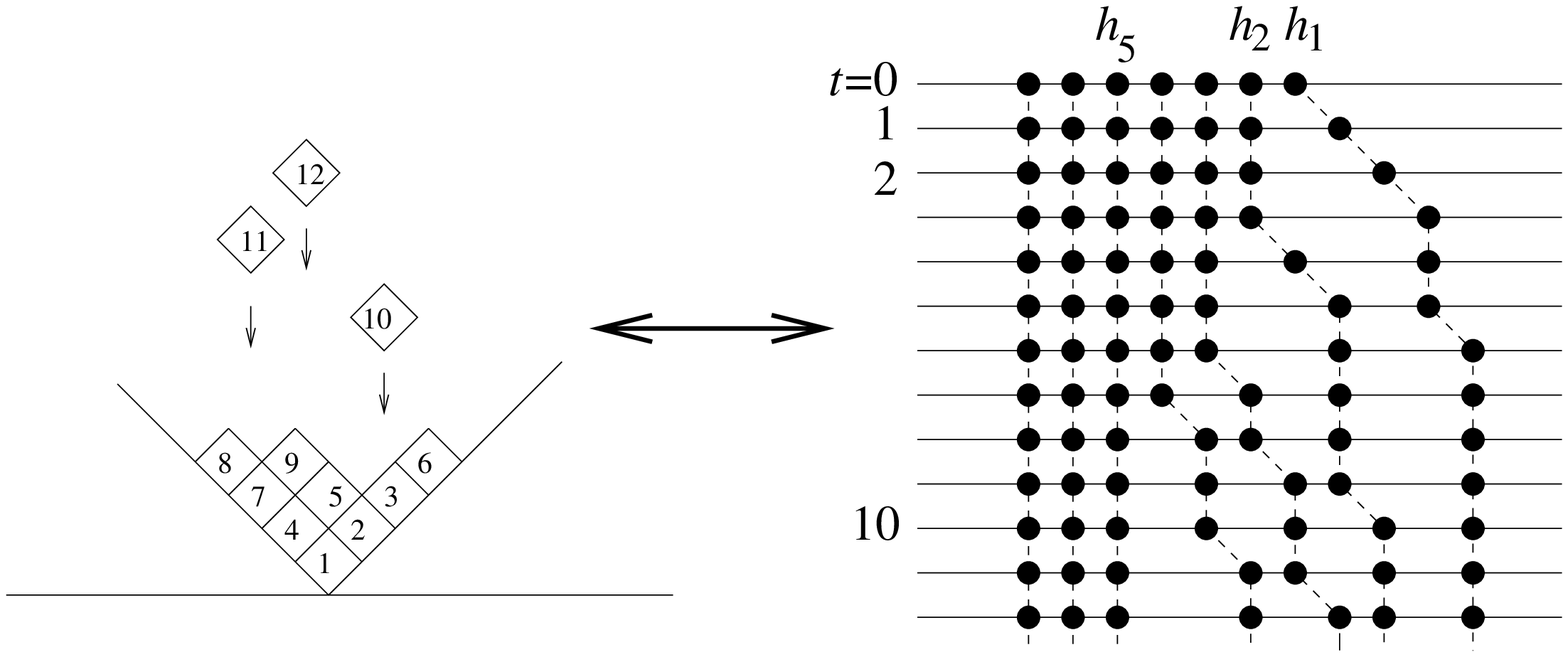}}}$$

Here, in this article, we give a method to compute the correlations of particle positions of the form:
\beq
\left< \left(\sum_{i=1}^N {1\over \xi_1-h_i}\right)\,\left(\sum_{i=1}^N {1\over \xi_2-h_i}\right)\, \dots \, \left(\sum_{i=1}^N {1\over \xi_k-h_i}\right) \right>
\eeq
for all $k$, and to any order in the large $q$ expansion.

\subsection{Algebraic geometry}

\subsubsection{Gromov-Witten theory of ${\mathbb P}^1$}

${\mathbb P}^1$ is the complex projective plane, i.e. the complex plane with a point at $\infty$, it is (the only) compact Riemann surface of genus $0$, it is also called the Riemann sphere.

A stable map $(C,p_1,\dots,p_n,f)$ to ${\mathbb P}^1$, is the data of a (possibly stable nodal) curve $C$ (of complex dimension $1$), with $n$ smooth marked points $p_1,\dots, p_n$, and a holomorphic function $f$ of degree $d$, from $C$ to ${\mathbb P}^1$.
A nodal curve, is a possibly degenerate compact Riemann surface (some cycles may have been pinched), and stability means that if we remove the pinched cycles, all connected domains have strictly negative Euler characteristics, so that there are only finitely many holomorphic automorphisms preserving the marked points.
The degree $d$ of map $f:C\to {\mathbb P}^1$, is the number of preimages of a generic point in ${\mathbb P}^1$. Critical points are the points with less than $d$ preimages, i.e. such that $df=0$.
Generically, if $C$ is a smooth curve, $f$ has $2d+2g-2$ critical points.

The set of all stable maps $(C,p_1,\dots,p_n,f)$ to ${\mathbb P}^1$,  is  a finite dimensional complex manifold (it is locally described by a finite number of complex parameters called moduli, indeed the data of a given degree function is more or less the data of the coefficients of a polynomial), which is called the moduli space.

Let $\overline{\cal M}_{g,n}({\mathbb P}^1,d)$ be the moduli space of stable maps of degree $d$ to ${\mathbb P}^1$.
Since each point $p_i$ is smooth, we have a natural line bundle ${\cal L}_i$ over $\overline{\cal M}_{g,n}({\mathbb P}^1,d)$, whose fiber is, for each point $(C,p_1,\dots,p_n,f)\in\overline{\cal M}_{g,n}({\mathbb P}^1,d)$, the cotangent space of $C$ at $p_i$, i.e. $T_{p_i}^* C$. We can consider the first Chern class $c_1({\cal L}_i)$ of this line bundle, i.e. the curvature form of a $U(1)$ connection.

There is a virtual fundamental homology class on $\overline{\cal M}_{g,n}({\mathbb P}^1,d)$, of dimension $2d+2g-2+n$ where the intersection theory can be computed (i.e. where classes $c_1({\cal L}_i)$ can be integrated).
By Poincarr\'e duality, classes are dual to cycles, and integrals of classes compute the number of intersection points of the cycles, provided that this number is finite, i.e. provided that the sum of dimensions is the total dimension of the whole space. If the dimension is wrong we say that the intersection number is zero. The virtual class, allows to extend the generic dimension of smooth curves $2d+2g-2$, to degenerate curves.

Also, some points of moduli spaces may have extra symmetries, and the moduli spaces are orbifolds (one needs to quotient by the symmetries), and intersection numbers are counted with the symmetry factor, so that they can be rational numbers instead of integers.

If we also fix the images of points $p_i\in C$ to be $q_i\in{\mathbb P}^1$, the Gromov--Witten invariants of ${\mathbb P}^1$ are defined as the intersection numbers \cite{witten1}:
\beq
<\tau_{k_1}\dots \tau_{k_m}>_d  
= \int_{[\overline{\cal M}_{g,n}({\mathbb P}^1,d)]^{\rm vir}\cap \{f(p_i)=q_i\}_{i=1\dots n}} \qquad c_1({\cal L}_1)^{k_1}\dots c_1({\cal L}_n)^{k_n} 
\eeq

Those intersection numbers count a finite number of intersection points in $\overline{\cal M}_{g,n}({\mathbb P}^1,d)$, and thus they count certain curves $C$ with marked points $p_1,\dots,p_n$ mapped to $q_1,\dots,q_n$. Those curves are instantons which extremize a topological sigma model action (type A topological string theory).
To construct a curve which passes through given points, it is sufficient to know the first few terms in the Taylor expansion near those points, and thus the degree of tangency at those points, i.e. the number of vanishing coefficients in the Taylor expansion. Instantons are entirely encoded by those degrees, which form a partition of the total degree $d$, and the intersection numbers can be rewritten in terms of sums over partitions.

In \cite{Okounkov4, Okounkov3}, it was proved that the Gromov-Witten invariants of ${\mathbb P}^1$ are:
\beq
<\tau_{k_1}\dots \tau_{k_m}>_d  = {1\over \prod_i (k_i+1)!}\,\sum_{\l,|\l|=d} {\cal P}(\l)\,\,\, \prod_i C_{k_i+1}(\l)  
\eeq

This sum is nearly of the type of what we computed in this article, we only have to take derivatives with respect to the $t_k$'s at $t_k=0$. The general method for computing the derivatives of $F_g$ with respect to any parameter of the spectral curve is explained in \cite{EOFg}, and can be applied here.
It would be interesting to 
study in further details, the consequences of our formula for Gromov-Witten invariants of ${\mathbb P}^1$.

\subsubsection{Gromov-Witten invariants of CY 3-folds and topological strings}
\label{secXp}

See \cite{nv, vonk, mmhouches, Behrend, Okounkov2bis, reBmodel} for an introduction to topological string theory and Toric Calabi-Yau 3-folds.

\bigskip

For any $p\in {\mathbb Z}$, consider the toric Calabi Yau 3-fold $X_p$, which was investigated by \cite{AgOSV, bryan, Marcosconj, CGMPS, forbes, forbes2}: 
\beq
O(p-2)\oplus O(-p) \,\, \longrightarrow {\mathbb P}^1
\eeq
which is a rank $2$ bundle over ${\mathbb P}^1$.
It is often represented by the toric fan diagram \cite{vonk}:

$$
\hbox{
\begin{picture}(0,40)(240,20)
\put(150,40){\line(1,0){30}}
\put(150,40){\line(-1,-1){20}}
\put(180,40){\line(1,-1){20}}
\put(150,40){\line(1,1){30}}
\put(180,40){\line(-1,1){13}}
\put(163,57){\line(-1,1){13}}
\end{picture}
}
$$
The partition function of the A-sigma model topological string theory in target space $X_p$, is defined as the generating function of Gromov--Witten invariants of $X_p$, i.e. a generating function for counting instantons, i.e. for computing intersection numbers:
\beq
\ln{\left(Z_{X_p}(g_s,t)\right)} = \sum_{g=0}^\infty g_s^{2g-2}\,F_g(t) = \sum_{g=0}^\infty \sum_{d=1}^\infty N_{g,d}(X_p)\,\ee{-dt}\,g_s^{2g-2}
\eeq
$N_{g,d}(X_p)$ are the Gromov Witten invariants of $X_p$, i.e. in some sense the number of instantons (rational curves embedded in $X_p$) of genus $g$ and degree $d$ (the numbers may be rational because of orbifold points).

It can be computed with the so-called Topological Vertex method \cite{topvertex}, and it can be written as a sum of partitions with Plancherel measure \cite{nekrasov, nekrasov2, Marcosconj, CGMPS}:
\beq\label{ZXp}
Z_{X_p}(g_s,t) = \sum_\l {\cal P}_q(\l)\,\,q^{(p-1)C_2(\l)/2} \,\ee{-t |\l|}
\virg
q=\ee{-g_s}
\eeq
where $|\l|=\sum_i \l_i$ is the weight of the representation $\l$, and $C_2(\l)=\sum_i \l_i(\l_i-2i+1)$ is the second Casimir, and ${\cal P}_q(\l)$ is the $q$-deformed Plancherel measure.

\medskip

Through mirror symmetry \cite{horivafa, Agvafa, bookmirror, forbes, forbes2}, the type A topological string theory on $X_p$, should be dual to a type B topological string theory on a mirror CY 3-fold, which we note $\td{X}_p$, and whose equation is expected to be of the form:
\beq
H(\ee{x},\ee{y}) = \xi \zeta
\eeq
(with $x,y,\xi,\zeta \in {\mathbb C}^4$, so that this is locally a 3-dimensional complex submanifold of ${\mathbb C}^4$), and where $H(u,v)$ is a polynomial in both variables.
Expressions of $H(u,v)$ (up to symplectic transformations) were derived in \cite{CGMPS, Marcosconj}.

\medskip

$F_0$ was computed in \cite{CGMPS, Marcosconj, forbes2}, and $F_1$ was computed in \cite{Marcosconj}. Marcos Mari\~ no conjectured that higher $F_g$'s with $g\geq 2$ are given 
by the symplectic invariants of \cite{EOFg} for the curve $H(\ee{x},\ee{y})=0$.

\smallskip
Here, we have proved that conjecture for $X_p$.

\medskip

To summarize, we have rewritten the type A model partition function of $X_p$ as a matrix integral, and therefore we obtained the topological small $g_s$ expansion from loop equations (in other words from Virasoro constraints). To leading order, the loop equations define a spectral curve (of mirror type).
Once the spectral curve is known, all the topological expansion of the matrix integral, and thus the Gromov-Witten invariants, are then obtained from the method of \cite{EOFg}, as was conjectured by
Marcos Mari\~ no \cite{Marcosconj}.

Finally, we check that the spectral curve of the matrix model (which coincides with the one found by \cite{CGMPS} is symplecticaly equivalent to the mirror curve $H(\ee{x},\ee{y})=0$, and therefore it indeed has the same $F_g$'s.

\section{Conclusion}

In this article, we have shown that the topological expansion of a sum over partitions with the Plancherel measure:
\beq
Z_N(q;t_k) = \sum_{n(\l)\leq N}\, {\cal P}(\l)\,\, q^{|\l|}\,\, \ee{-\sqrt{q}\sum_k {t_k\, q^{-k/2}\over k} \, C_k(\l)}
\eeq
is a matrix integral, and has a topological expansion of the form:
\beq
\ln{Z_N(q;t_k)} \sim \sum_{g=0}^\infty q^{1-g}\,\, F_g(t_k)
\eeq
where the $F_g$'s are the symplectic invariants of the spectral curve:
\beq
\left\{\begin{array}{l}
\displaystyle x(z) = \ee{-u_0}\,(z+{1\over z}-u_1) \cr \cr
y(z) = \ln{(z)}+ {1\over 2}\,\sum_{k=1}^d u_k (z^k-z^{-k})  
\end{array}\right.
\eeq
Moreover, the same method also gives the large $q$ asymptotic expansion, to all orders, of the density correlation functions.

Our results were derived in a 1-cut regime, but it seems easy to extend them to any other regime, the only difference is that we would need $\theta$-functions instead of the Zhukovski rational parametrization.

\bigskip

We have also obtained similar results for the $q-$deformed Plancherel measure:
\beq
Z_N(q;t_k) = \sum_{n(\l)\leq N}\, {\cal P}_q(\l)\,\,  \ee{-{1\over g_s}\,\sum_k {t_k\, g_s^{k}\over k} \, C_k(\l)}
\virg q=\ee{-g_s}
\eeq
which is also a matrix model, whose topological expansion is of the form:
\beq
\ln{Z_N} \sim \sum_{g=0}^\infty g_s^{2g-2}\,\, F_g(t_k)
\eeq
When $t_1=t, t_2=p-1$, the sum $Z_\infty$ is the generating function of Gromov-Witten invariants of the calabi-Yau 3-fold $X_p$. The spectral curve is:
\beq
\left\{
\begin{array}{l}
x(z)  =  {(1-{z\over z_0})(1-{1\over z z_0})\over (1+{1\over z_0})^2}  \cr
y(z) = {1\over x(z)}\,\left(- \ln{z}  + {p\over 2}\,\ln{\left(1-z/z_0\over 1-1/z z_0\right)}  \right) 
\end{array}\right.
\virg \ee{-t} = {1\over z_0^2}\,\,(1-{1\over z_0^2})^{p(p-2)}
\eeq

We have thus proved that $Z_N$ can be obtained as a matrix integral (a unitary matrix integral, because the eigenvalues are constrained on a circle), and that the $F_g$'s are the symplectic invariants of the mirror spectral curve $H(\ee{x},\ee{y})=0$.
This proves the conjecture of M. Mari\~ no \cite{Marcosconj} for $X_p$.

Dijkgraaf and Vafa recentely noticed that that the symplectic invariants of \cite{EOFg} are also obtained from the Kodaira--Spencer theory  \cite{DVKS}, and thus we have proved that the type B topological string theory on the mirror of ${X}_p$, is equivalent to a matrix model, and to the Kodaira--Spencer theory.

\medskip

To go further in the same direction for all toric Calabi-Yau 3-folds, one would need to extend the method to plane partitions, because the topological vertex \cite{topvertex} is expressed as a sum of plane partitions, and it is only for $X_p$ that it reduces to usual partitions. Then one needs to "glue" topological vertices together.
This is a work in progress \cite{eyntopvertex}, but with more involved tools because the spectral curve is no longer hyperellitical.

\bigskip

Many other extensions of the present method can be explored.
First, we have not studied in details the consequences of our method for crystal growth, longest increasing sequence, or T.A.S.E.P..

Also, we have not studied in detail the expectation values of Casimirs. That could have interesting conseqences in the Hurwitz numbers generating functions.
Finaly, we have not computed the spectral curve of the $q-$defomed Plancherel measure, with higher Casimirs. The method works in the same way, but the solution is slightly more difficult to write.
It would be interesting to continue this computation.

The same kind of sums over partitions also appears in Seiberg-Witten theory, as discussed in \cite{nekrasov3}. We indeed find the same spectral curve and same prepotential $F_0$. The authors of \cite{nekrasov3} found their results from the integrable hierarchy structure, and indeed the symplectic invariants of \cite{EOFg} have such an integrable structure (see section 9 in \cite{EOFg}). It would be interesting to explore further the computations for that case.

\bigskip

Also, we have studied only 1-cut cases, which are the most relevant for combinatorics formal generating series.
However, for many applications to statistical physics, we have convergent series, and it would be interesting to study phase transitions. When there are several cuts, the matrix integral no longer has a topological expansion, it receives an extra oscillatory part, as dicussed in \cite{BDE, eynff}.

\bigskip

\section*{Acknowledgments}
We would like to thank Marcos Mari\~ no for numerous discussions and encouragements on this subject, as well as careful reading and comments on the manuscript.
We would also like to thank A. Alexandrov, P. Di Francesco, R. Dijkgraaf, P. Ferrari, K. Mallick, A. Okounkov, N. Orantin, and N. Nekrasov, for useful and fruitful discussions on this subject.
This work is partly supported by the Enigma European network MRT-CT-2004-5652, by the ANR project G\'eom\'etrie et int\'egrabilit\'e en physique math\'ematique ANR-05-BLAN-0029-01, by the Enrage European network MRTN-CT-2004-005616,
by the European Science Foundation through the Misgam program,
by the French and Japaneese governments through PAI Sakurav, by the Quebec government with the FQRNT.


\begin{thebibliography}{99}
\bibliographystyle{plain}

\bibitem{Aldous} D.J. Aldous, P. Diaconis, Longest increasing subsequences: from patience sorting to the Baik-Deift-Johansson theorem, Bull. Amer. Math. Soc. 36 (1999), 413-432.

\bibitem{AgOSV} M. Aganagic, H. Ooguri, N. Saulina, C. Vafa, Black Holes, q-Deformed 2d Yang-Mills, and Non-perturbative Topological Strings, Nucl.Phys. B715 (2005) 304-348, hep-th/0411280.

\bibitem{topvertex} M. Aganagic, A. Klemm, M. Mari\~ no, C. Vafa, The topological vertex, hep-th 0305132.

\bibitem{Agvafa} M. Aganagic and C. Vafa, ÒMirror symmetry, D-branes and counting holomorphic 
discs,Ó hep-th/0012041. 


\bibitem{baik} J. Baik, E.M. Rains, Symmetrized random permutations, Random Matrix Models and Their Applications, vol. 40, Cambridge University Press, 2001, pp. 1-19.

\bibitem{BDJ} J. Baik, P. Deift, K. Johansson, On the distribution of the length of 
the longest increasing subsequence of random permutations, Journal of 
AMS, 12 (1999), 1119Ð1178.

\bibitem{Behrend} K. Behrend, Gromov-Witten invariants in algebraic geometry, Invent. 
Math. 127 (1997), 601Ð617. 

\bibitem{BDE} G. Bonnet, F. David, B. Eynard, Breakdown of universality in multicut matrix models, J.Phys.A33:6739-6768,2000,
e-Print: cond-mat/0003324.

\bibitem{reBmodel} V. Bouchard, A. Klemm, M. Mari\~ no, S. Pasquetti, Remodelling the B-model, hep-th 0709.1453.

\bibitem{BIPZ}
E.~Br\'ezin, C.~Itzykson, G.~Parisi and J.~B.~Zuber,
``Planar Diagrams,''
Commun.\ Math.\ Phys.\  {\bf 59}, 35 (1978).

\bibitem{bryan} J. Bryan and R. Pandharipande, The local Gromov-Witten theory of curves. 
math.AG/0411037. 

\bibitem{CGMPS} N. Caporaso, L. Griguolo, M. Mari\~ no, S. Pasquetti, D. Seminara, Phase transitions,  doubleÐscaling limit,  and topological strings, hep-th 0606120.


\bibitem{CE} L. Chekhov and B. Eynard, ÒHermitean matrix model free energy: Feynman graph 
technique for all genera,Ó JHEP 0603, 014 (2006) [arXiv:hep-th/0504116]. 

\bibitem{chiral} M. J. Crescimanno and W. Taylor, ÒLarge N phases of chiral QCD in two- 
dimensions,Ó Nucl. Phys. B 437, 3 (1995) [arXiv:hep-th/9408115]. 

\bibitem{derrida} Derrida, An exactly soluble non-equilibrium system: the asymmetric simple exclusion process, Phys. Rep. 301, 65, (1998).

\bibitem{derrida2} B. Derrida, M. R. Evans, V. Hakim, V. Pasquier, 1993, Exact solution of a 1D asymmetric exclusion model using a matrix formulation, J. Phys. A: Math. Gen. 26, 1493.

    
\bibitem{DVKS} R. Dijkgraaf, C. Vafa, Two Dimensional Kodaira-Spencer Theory and Three Dimensional Chern-Simons Gravity, arXiv:0711.1932 [hep-th].
  
\bibitem{eynloop1mat}
 B.~Eynard,
``Topological expansion for the 1-hermitian matrix model correlation
functions,''
arXiv:hep-th/0407261.
 
 \bibitem{eynform}
 B.~Eynard, ``Formal matrix integrals and combinatorics of maps,''
  arXiv:math-ph/0611087.
 
  
\bibitem{EOFg}
B. Eynard and N. Orantin, ``Invariants of algebraic curves and topological expansion'',
arXiv:math-ph/0702045.

\bibitem{EOsym} B. Eynard and N. Orantin, Topological expansion of mixed correlations in the hermitian 2 Matrix Model and $x-y$ symmetry of the $F_g$ algebraic invariants, math-ph/arXiv:0705.0958, to appear in J.Phys A.

\bibitem{eyntopvertex} B. Eynard, private notes on matrix integrals for counting plane partitions. In progress.

\bibitem{eynbirthcut} B. Eynard, Universal distribution of random matrix eigenvalues near the ``birth of a cut'' transition, math-ph/0605064, JSTAT, Theory and Experiment P07005 (2006) P07005.

\bibitem{eynff} B. Eynard, Large N expansion of convergent matrix integrals, holomorphic anomalies, and background independence, math-ph: arXiv.0802.1788.

\bibitem{ferrari1}  P. L. Ferrari, M. Praehofer, One-dimensional stochastic growth and Gaussian ensembles of random matrices,
 proceedings of ''Inhomogeneous Random Systems 2005'', Markov Processes Relat. Fields 12 (2006) 203-234, arXiv:math-ph/0505038.

 \bibitem{ferrari2} Patrik L. Ferrari, Polynuclear growth on a flat substrate and edge scaling of GOE eigenvalues, Comm. Math. Phys., 252 (2004), 77-109,     arXiv:math-ph/0402053.

\bibitem{forbes} B. Forbes and M. Jinzenji, ÒJ functions, non-nef toric varieties and equivariant local 
mirror symmetry of curves,Ó arXiv:math.ag/0603728. 

\bibitem{forbes2} B. Forbes and M. Jinzenji, ÒLocal mirror symmetry of 
curves: Yukawa couplings and genus 1,Ó arXiv:math.ag/0609016. 


\bibitem{Mallick} O. Golinelli, K. Mallick, 2006, The asymmetric simple exclusion process : an integrable model for non-equilibrium statistical mechanics, J. Phys. A: Math. Gen. 39 12679.
  
\bibitem{horivafa} K. Hori and C. Vafa,ÒMirror symmetry,Ó hep-th/0002222.
 
 \bibitem{bookmirror} K. Hori, S. Katz, A. Klemm, R. Pandharipande, R. Thomas, C. Vafa, R. Vakil, E. 
Zaslow,   AMS, Providence, 2003. 

\bibitem{Jarctic} K. Johansson, The arctic circle boundary and the Airy process, Ann. Probab. 33 (2005), 1-30.

\bibitem{johansson} K. Johansson, The longest increasing subsequnce in a random permuta- 
tion and a unitary random matrix model, Math. Res. Lett., 5 (1998), 
no. 1-2, 63Ð82.

\bibitem{kerov1} S. Kerov, Gaussian limit for the Plancherel measure of the symmetric 
group, C. R. Acad. Sci. Paris, 316, S?erie I, 1993, 303Ð308.

\bibitem{kerov2} S. Kerov, Transition probabilities of continual Young diagrams and the 
Markov moment problem, Func. Anal. Appl., 27, 1993, 104Ð117.

\bibitem{kerov3} S. Kerov, A differential model of growth of Young diagrams, Proceedings 
of the St. Petersburg Math. Soc., 4, 1996, 167Ð194. 


\bibitem{KSW} I. K. Kostov, M. Staudacher and T. Wynter, ÒComplex matrix models and statis- 
tics of branched coverings of 2D surfaces,Ó Commun. Math. Phys. 191, 283 (1998) 
[arXiv:hep-th/9703189]. 

\bibitem{liggett} T.M. Liggett, Coupling the simple exclusion process, Ann. Probab. 4 (1976), 339-356.

\bibitem{Logan} B. F. Logan and L. A. Shepp, A variational problem for random Young 
tableaux, Adv. Math., 26, 1977, 206Ð222.

\bibitem{mmhouches}
  M.~Mari\~no, ``Les Houches lectures on matrix models and topological strings,''
  arXiv:hep-th/0410165.

\bibitem{Marcosconj} M. Mari\~ no, Open string amplitudes and large order behavior  
in topological string theory, hep-th 0612127.

\bibitem{nekrasov3}  A. Marshakov, N. Nekrasov, Extended Seiberg-Witten Theory and integrable hierarchy, hep-th/0612019.

 \bibitem{nv}
 A.~Neitzke and C.~Vafa, 
 ``Topological strings and their physical applications,''
  arXiv:hep-th/0410178.

\bibitem{nekrasov}
N.~A.~Nekrasov, ``Seiberg-Witten prepotential from instanton counting,''
  Adv.\ Theor.\ Math.\ Phys.\  {\bf 7}, 831 (2004)
  [arXiv:hep-th/0206161].

\bibitem{nekrasov2}  N. Nekrasov, A. Okounkov, Seiberg-Witten Theory and Random Partitions, hep-th/0306238.

\bibitem{Okounkov1} A. Okounkov, ``the uses of random partitions'', math-ph/0309015.

\bibitem{Okounkov2} A. Okounkov, Asymptotics of Plancherel measures for symmetric groups, Journal of American Mathematical Society, Vol 13, 3, 481-515.

\bibitem{Okounkov2bis} A. Okounkov,  Random surfaces enumerating algebraic curves, hep-th 0412008.

\bibitem{Okounkov3} A. Okounkov and R. Pandharipande, Gromov-Witten theory, Hurwitz 
numbers, and matrix models, I, math.AG/0101147. 

\bibitem{Okounkov4} A. Okounkov and R. Pandharipande, Gromov-Witten theory, Hurwitz 
theory, and completed cycles, math.AG/0204305.

\bibitem{spohn} M. Pr\"ahofer and H. Spohn, Universal Distributions for Growth Processes 
in 1+1 Dimensions and Random Matrices, Phys. Rev. Lett. 84, 4882 
(2000). 

\bibitem{Robinson} G. de B. Robinson, "On representations of the symmetric group," Amer. J. Math. 60 (1938), 745Ð760.

\bibitem{ssuites1} C. Schensted, Longest increasing and decreasing subsequences, Canad. J. Math., 13, 1961,  179-191. MR 22:12047 

\bibitem{thooft}
G.~'t Hooft, ``A planar diagram theory for strong interactions,''
Nucl.\ Phys.\ B {\bf 72} (1974) 461.

\bibitem{Planchrel1} A. Vershik and S. Kerov, Asymptotics of the Plancherel measure of the symmetric group and  the limit form of Young tableaux, Soviet Math. Dokl., 18, 1977, 527Ð531. 

\bibitem{vershik} A. Vershik, Statistical mechanics of combinatorial partitions and their 
limit configurations, Func. Anal. Appl., 30, no. 2, 1996, 90Ð105.

\bibitem{vonk} M. Vonk, ÒA mini-course on topological strings,Ó arXiv:hep-th/0504147. 

\bibitem{TW} C. A. Tracy and H. Widom, Level-spacing distributions and the Airy 
kernel, Commun. Math. Phys., 159, 1994, 151Ð174.


\bibitem{witten1} E. Witten, ÔTopological sigma models,Ó Commun. Math. Phys. 118, 411 (1988). ÒOn 
the structure of the topological phase of two-dimensional gravity,Ó Nucl. Phys. B 340, 
281 (1990). 



\end{thebibliography}
\end{document}